\documentclass[lettersize,journal]{IEEEtran}
\usepackage{amsmath,amsfonts}
\usepackage{algorithmic}
\usepackage{algorithm}
\usepackage{array}
\usepackage[caption=false,font=normalsize]{subfig}
\usepackage{textcomp}
\usepackage{stfloats}
\usepackage{url}
\usepackage{verbatim}
\usepackage{graphicx}
\usepackage{cite}

\usepackage{subfig}

\usepackage{multirow}
\usepackage{booktabs}
\usepackage{subcaption}
\usepackage{float}
\usepackage{makecell}
\usepackage{enumitem}
\usepackage{listings}
\usepackage{pifont}
\usepackage{utfsym}
\usepackage{soul} 
\usepackage{subfloat}
\usepackage{filecontents}
\usepackage[colorlinks=true, linkcolor=blue, citecolor=blue, urlcolor=blue]{hyperref}

\begin{document}

\title{\textit{I Know Who Clones Your Code:} Interpretable Smart Contract Similarity Detection}

% \author{\IEEEauthorblockN{Anonymous}}
\author{Zhenguang Liu, Lixun Ma, Zhongzheng Mu, Chengkun Wei, Xiaojun Xu, Yingying Jiao, Kui Ren,~\IEEEmembership{Fellow,~IEEE}
        % <-this % stops a space
\thanks{
This work is supported by the National Key R\&D Program of China (No. 2023YFB3105904), the Key R\&D Program of Zhejiang Province (No. 2023C01217), and the National Natural Science Foundation of China (No. 62372402). \textit{(Corresponding authors: Yingying Jiao; Lixun Ma; Zhongzheng Mu.)}} 
\thanks{
Zhenguang Liu, Kui Ren, Chengkun Wei and  Lixun Ma are with Zhejiang University (The State Key Laboratory of Blockchain and Data Security, Zhejiang University), (e-mail: liuzhenguang2008@gmail.com; kuiren@zju.edu.cn; weichengkun@zju.edu.cn; malx@zju.edu.cn). 

Zhenguang Liu is also with Shandong Rendui Network Co., Ltd., and is with Hangzhou High-Tech Zone (Binjiang) Institute of Blockchain and Data Security. 

Zhongzheng Mu and Xiaojun Xu are with the School of Computer and Information Engineering, Zhejiang Gongshang University, China (e-mail: 22020100068@pop.zjgsu.edu.cn; xuxj2022@gmail.com).

Yingying Jiao is with Zhejiang University of Technology, Hangzhou 310023, China (e-mail: yingyingjiao21@gmail.com).

} 
% \thanks{

% Corresponding to: Yingying Jiao, Lixun Ma, Zhongzheng Mu (e-mail: yingyingjiao21@gmail.com; malx@zju.edu.cn; 22020100068@pop.zjgsu.edu.cn).
% }
}

\markboth{IEEE TRANSACTIONS ON DEPENDABLE AND SECURE COMPUTING}%
{Liu \MakeLowercase{\textit{et al.}}: \bf \textit{I Know Who Clones Your Code}: Interpretable Smart Contract Similarity Detection}


\maketitle

\begin{abstract}
  Widespread reuse of open-source code in smart contract development boosts programming efficiency but significantly amplifies bug propagation across contracts, while dedicated methods for detecting similar smart contract functions remain very limited. 
  Conventional abstract-syntax-tree (AST) based methods for smart contract similarity detection face challenges in handling intricate tree structures, which impedes detailed semantic comparison of code. 
  Recent deep-learning based approaches tend to overlook code syntax and detection interpretability, resulting in suboptimal performance.
  
  To fill this research gap, we introduce \textsc{SmartDetector}, a novel approach for computing similarity between smart contract functions, explainable at the fine-grained statement level. 
  Technically, \textsc{SmartDetector} decomposes the AST of a smart contract function into a series of smaller statement trees, each reflecting a structural element of the source code. 
  Then, \textsc{SmartDetector} uses a classifier to compute the similarity score of two functions by comparing each pair of their statement trees. 
  To address the infinite hyperparameter space of the classifier, we mathematically derive a cosine-wise diffusion process to efficiently search optimal hyperparameters. 
  Extensive experiments conducted on three large real-world datasets demonstrate that \textsc{SmartDetector} outperforms current state-of-the-art methods by an average improvement of 14.01\% in F1-score, achieving an overall average F1-score of 95.88\%.
\end{abstract}

\begin{IEEEkeywords}
Smart contract, blockchain, code clone detection. 
\end{IEEEkeywords}

\section{Introduction}
\IEEEPARstart{S}{mart} contract~\cite{zou2019smart} has risen as the hallmark application of 
blockchain. 
Essentially, smart contracts are code running on top of a blockchain system, which 
formulates contract terms into executable blockchain programs~\cite{liu2021combining}. 
Due to the \emph{immutability} and \emph{multi-parties certification} nature of 
blockchain, the contract terms will be rigorously and automatically followed during 
execution, ensuring fairness for all parties involved in the contract. 
So far, over 10 million smart contracts are deployed on the Ethereum 
mainnet~\cite{he2020characterizing}, which is one of the most renowned blockchains. 
Surprisingly, these smart contracts currently hold virtual currencies valued at 
over 397.8 million US dollars~\cite{smartcontractvalue}, and the number of assets 
controlled by smart contracts continues to grow rapidly each year. 

To improve programming efficiency, smart contract developers frequently clone 
code snippets from other open-source contracts to implement similar functionalities. 
This, unfortunately, also leads to two serious practical issues. 
First, code replication exposes developers to the risks of integrating vulnerabilities and attacker-placed malicious code into contracts. 
A notable instance is the famous ``honeypot'' smart contract, which is a scam contract designed to defraud users~\cite{he2020characterizing} but has been copied many times. 
\textcolor{black}{Second, code cloning can also potentially violate copyright laws, especially when the original code is copyrighted or protected by a license agreement~\cite{luo2014semantics}. For example, \textit{Compuware} filed an intellectual property lawsuit against \textit{IBM}, which resulted in \textit{IBM} paying \$140 million in licensing fees and an additional \$260 million to acquire \textit{Compuware}’s services~\cite{compuware_ibm_2013} 
While there have been no major incidents of this nature in the smart contract domain thus far, relevant legal frameworks are continuously evolving. This will undoubtedly become an important issue to address in the future.}

\begin{figure}[t] 
  \centering 
  \setlength{\abovecaptionskip}{0.4cm} 
  \includegraphics[width=.8\linewidth]{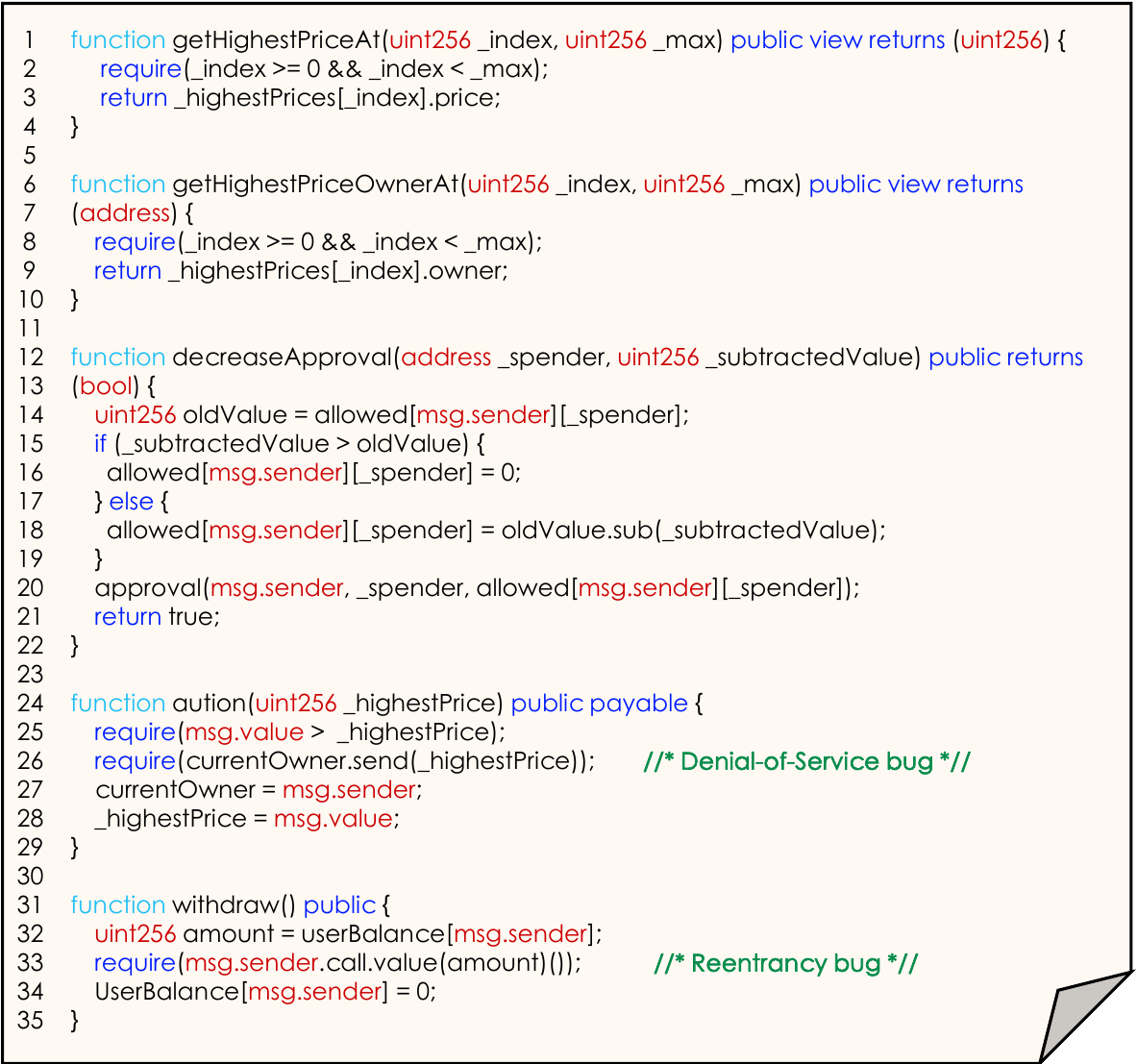} 
  \caption{\textbf{Five most frequently reused templates. } }
  \label{fivefunctions} 
\end{figure}

Interestingly, we empirically identified the five most prevalent function templates 
on Ethereum, each of which is copied and reused more than 2,000 times in other smart 
contract functions. 
We found that three of the five functions are benign while the other two have vulnerabilities. 
The five functions are illustrated in Figure~\ref{fivefunctions}. 
Among them, the {\texttt{auction()}} function harbors a denial-of-service vulnerability, 
which leads to the KotET (The King of the Ether Throne) illegal prize winner 
event~\cite{KoET}. 
Surprisingly, the {\texttt{auction()}} function is copied in other 2,083 functions, 
causing serious vulnerability propagation. 
Another function {\texttt{withdraw()}} is reused in other 2,022 functions, which is 
vulnerable to reentrancy attacks, resulting in the infamous DAO incident and causing 
$\$60$ million losses~\cite{theDAO}. 

\begin{figure*}[t] 
  \centering 
  \setlength{\abovecaptionskip}{0.4cm} 
  \includegraphics[width=.8\linewidth]{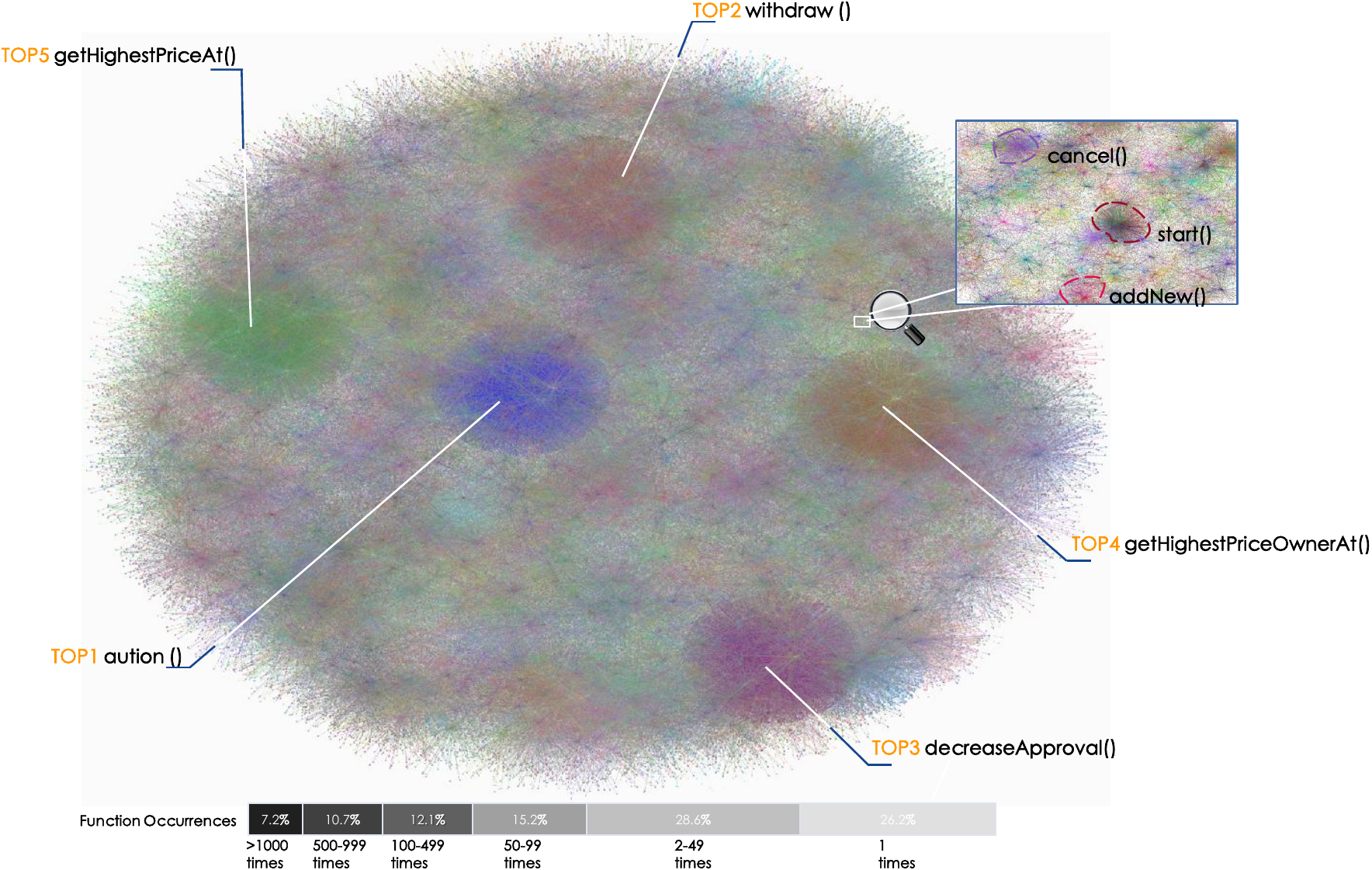} 
  \caption{\textbf{Panorama of Ethereum code cloning phenomenon. }}  
  \label{para} 
\end{figure*}

\textbf{Existing works} for smart-contract function similarity detection can be roughly 
cast into two categories. 
\textit{One line of works} concentrate on designing hand-crafted code patterns for function 
similarity computation. 
For instance, Volcano~\cite{samreen2022volcano} designs different types of code patterns 
to detect smart contract clones. 
Eclone~\cite{eclone} proposes smart contract birthmarks that sketch symbolic 
execution traces and maintain syntactic regularities for function similarity detection.  
\textit{Another line of works} resort to abstract syntax trees (AST), which 
capture rich information about the smart contract code structure and syntax. 
SmartEmbed~\cite{smartembed} serializes and encodes the nodes of AST into a 
stream of tokens, stacking them into a matrix for similarity detection. 
SRCL~\cite{SRCL} learns local and global information from AST by utilizing 
 Transformer and CNN encoders. 

Upon investigating and experimenting with the released code of state-of-the-art 
methods~\cite{smartembed}~\cite{SRCL}~\cite{eclone}, we 
empirically observe that two issues still persist: 
(\textbf{1}) Hand-crafted code patterns are often too rigid to capture 
complex and global code semantics, leading to a high false negative rate. 
For AST (abstract syntax tree) based methods, 
while ASTs preserve code syntactic and semantic 
information~\cite{maletic2002source}, not all elements contribute equally to similarity detection. 
For example, \textit{variable names}, often arbitrarily 
assigned by users, are less critical for clone detection. 
Current AST-based methods, 
however, generally fail to discriminate between 
useful and non-essential information, adversely affecting performance. 
(\textbf{2}) The effectiveness of most existing methods heavily rely on the selection of classifier hyperparameters. 
Inappropriate hyperparameter settings lead to poor model performance. 
The vast and high-dimensional hyperparameter space poses a challenge to 
efficiently search for the optimal configuration. 

Motivated by this, we introduce \textsc{SmartDetector}, an automated and 
interpretable approach for smart contract function similarity detection. 
Technically, we embrace three key designs: \textbf{(i)} we propose to decompose the large 
abstract syntax tree (AST) of a function into smaller units, which we term as 
\emph{statement trees}. 
This alleviates the burden of maintaining and extracting 
information from the large AST, and facilitates code semantic comparison 
at the base (statement) level.  
\textbf{(ii)} We propose a hyperparameter evaluation neural network and a diffusion 
probabilistic model~\cite{ho2020denoising} to search for the optimal hyperparameters
of the classifier. 
We theoretically derived the hyperparameter sampling chain process for the diffusion probabilistic 
model. 
\textbf{(iii)} Since different types of semantic nodes in the statement tree have varying degrees of importance in similarity detection, 
we extract and categorize semantic nodes of a statement tree by considering its control flow statements, operators, variables, and used functions. 

\noindent\textbf{Experiments.}
We conducted extensive experiments on three large real-world datasets, covering 
over $3.28\times 10^5$ smart contract function pairs and $1.79\times 10^6$ 
statement tree pairs. 
Empirical results demonstrate that our method significantly outperforms existing methods, achieving a promising average F1-score of 95.88\% and a significant average improvement of 14.01\% in F1-score over current state-of-the-art approaches.

Interestingly, for a more profound comprehension of the smart contract code cloning phenomenon, 
we have created, to the best of our knowledge, the \textit{first} function similarity 
panorama of all Ethereum open-sourced functions in Figure~\ref{para}. 
In the figure, each node represents a real-world smart contract function. 
In total, there are 347,418 functions, which are partitioned into 14,345 
groups according to their similarities computed by our model.
Each function is connected to a representative function of its group through an edge, 
with the representative function termed as the template function. 
All nodes in the same group, as well as the edges connecting them, are drawn in 
the same color, while nodes from distinct groups are depicted in different colors. 
We also identify and highlight the five most frequently occurring template functions in the 
figure. 

To facilitate further research in this area, we have made our code, feature exactor, hyperparameter optimizer, and datasets publicly available at \url{https://github.com/alllgi/SmartDetecter}. 
To summarize, our contributions are: 

\begin{itemize}
  \setlength{\itemsep}{0pt}
  \setlength{\parsep}{0pt}
  \setlength{\parskip}{0pt}
      \item We propose \textsc{SmartDetector}, an interpretable method for detecting function-level similarity in smart contracts. 
      \textsc{SmartDetector} explicitly localizes the line numbers of similar code. 
      It introduces a novel strategy to decompose the entire AST into statement trees based on code execution logic, and captures both code features and structural-level syntactical knowledge for similarity detection. 
      \item We present a diffusion probabilistic model~\cite{ho2020denoising} and a hyperparameter evaluation neural network to search for the optimal hyperparameters of the classifier. 
      We mathematically derive the chain process for sampling classifier hyperparameters. 
      We also evaluate the importance of different features extracted from the 
      statement trees, and assign distinct weights to different features. 
      \item  \textsc{SmartDetector} was rigorously tested on over 1.79 million 
      statement tree pairs from 347,418 real-world smart contract functions. 
      Extensive empirical results show that \textsc{SmartDetector} achieves a 14.01\% average improvement in F1-score over current state-of-the-art methods, obtaining an impressive average F1-score of 95.88\%. 
      As a side contribution, we have released our implementation and dataset, hoping to 
      inspire future research. 
\end{itemize}

\noindent\textbf{Roadmap.}
This paper is structured as follows: 
Sections~\ref{Relatedwork} provides a review of related work. 
Section~\ref{Motivating Examples} elucidates the motivation behind our research. 
Section~\ref{Method Overview} describes the problem we aim to address and the overall workflow of \textsc{SmartDetector}. 
Sections~\ref{Design of statement tree}, ~\ref{Hyperparameters Optimization}, and ~\ref{Feature Extraction and Similarity Computation} introduce the details of our approach. 
Section~\ref{Evaluate} discusses the results of our evaluation. 
Finally, Sections~\ref{Conclusion} present the concluding remarks.

%---------------------------------------Related work----------------------------------------------------------------------------
\section{Related work} \label{Relatedwork}
\subsection{Code Clone Detection for Traditional Language}
Code clone detection is widely recognized as one of the most critical issues in the field of software engineering. 
Code clone detection for traditional language has consistently attracted significant attention due to its substantial impact on code quality, maintainability, and overall software development efficiency. 
Early works~\cite{golubev2021multi}~\cite{gode2009incremental}~\cite{li2017cclearner}~\cite{wang2018ccaligner} predominantly targeted syntactic similarities, which treat code fragments as natural language texts and model them based on token sequences. 
For example, \textit{CCFinder}~\cite{kamiya2002ccfinder} introduces a suffix-tree matching algorithm to identify duplicated token subsequences.  
\textit{CP-Miner}~\cite{li2006cp} applies data mining techniques for copy-paste detection. 
Thereafter, \textit{SourcererCC}~\cite{sajnani2016sourcerercc} utilizes token ordering combined with an optimized inverted-index technique. 
However, syntactic similarity detection often relies on the surface structure of the code, overlooking its underlying logic and intent. 
While these methods achieve commendable results, they struggle to effectively capture semantic similarities. 

Recent efforts have shifted towards detecting similar code by engaging in AST and graph analysis. 
One stream of works~\cite{chen2014achieving}~\cite{krinke2001identifying}~\cite{wang2017ccsharp}~\cite{zou2020ccgraph} represent the semantic information of code as graph structures, which are then learned by graph neural networks to detect code similarity. 
\cite{komondoor2001using} is one of the pioneering works that employ program dependence graphs (PDG) to identify code duplicates based on isomorphic PDG subgraphs. 
\textit{Gemini}~\cite{gemini} represents the code as a low-dimensional vector using a manually designed feature extraction method, and trains a Siamese neural network that calculates the similarities. 
However, this compression process may result in the loss of significant code semantic information. 
\textit{DeepSim}~\cite{deepsim} represents variables and basic code blocks, along with their relationships, as a binary matrix. 
This matrix is then fed into a deep learning model to detect similarities. 
\textit{FCCA}~\cite{FCCA} extracts code representations by integrating both unstructured forms (\textit{e.g.}, tokens) and structured forms (\textit{e.g.}, control flow graphs). 
These representations are then processed by a model with an attention mechanism to identify similarities. 

Another stream of works~\cite{hu2022treecen}~\cite{koschke2006clone}~\cite{wahler2004clone}~\cite{wei2017supervised}~\cite{wu2022detecting}~\cite{zhang2019novel} resort to using tree-based approaches for code clone detection. 
These methods typically involve extracting the abstract syntax tree (AST) from code, followed by structural and syntactic analysis to identify potential code clones. 
Early work \textit{CloneDR}~\cite{baxter1998clone} constructs abstract syntax trees (ASTs) of code, hashes each subtree and detects clones by comparing the resulting hash codes. 
Gabel et al.~\cite{gabel2008scalable} extract related structured syntax from PDG subgraphs, then convert it into tree structures. 
\textit{Deckard}~\cite{deckard} represents source code as numerical vectors derived from abstract syntax trees and uses clustering to identify similar ASTs. 
However, using clustering for similarity computation may not effectively distinguish between subtle differences in code. 
\textit{Code2Vec}~\cite{code2vec} encodes abstract syntax tree paths into fixed-length vectors to represent code semantic and predict code similarities. 
Nevertheless, these fixed-length vectors struggle to accurately capture the semantic meaning of the code.

\subsection{Smart Contract Similarity Detection}
Given the immutable nature of smart contracts, patching their vulnerabilities is nearly impossible, regardless of the amount of money the contract holds or its popularity. 
The widespread code cloning among smart contracts can easily propagate vulnerabilities, leading to systemic risks and financial losses. 
Consequently, smart contract similarity detection has become critical for blockchain security. 

One group of works focus on designing hand-crafted code patterns for functions similarity computation. 
For instance, \textit{Volcano}~\cite{samreen2022volcano} devises various code patterns to identify contract clones using signature matching. 
However, human-crafted code patterns are often too rigid and fail to capture complex global code semantics. 
Graph-based methods leverage Program Dependency Graph (PDG) or Control Flow Graph (CFG) to obtain comprehensive code semantic information, thereby achieving good detection results. 
\textit{Eclone}~\cite{eclone} captures the high-level semantics of a smart contract through symbolic transaction sketches and integrates this information with other syntactic data for similarity computation. 

Another group of works resort to the abstract syntax trees (AST), which capture detailed information about the structure and syntax of smart contract code. 
\textit{SmartEmbed}~\cite{smartembed} focuses on converting all AST elements into numerical vectors, which are then organized into a matrix for code similarity detection. 
\textit{SRCL}~\cite{SRCL} converts ASTs into pairs of type and value to capture both global and local semantic information, thereby considering a more comprehensive range of semantic information. 

Several other studies have explored the use of large language models for detecting smart contract clone behavior. 
For instance, \textit{ZC3}\cite{zc3} employs the large language model \textit{CodeBERT}~\cite{feng2020codebert}, combined with domain-aware learning and cycle consistency learning. 
However, compared to tree-based and graph-based methods, large language model-based approaches face the inherent challenge of effectively capturing the structural information and syntactic features of code.

%---------------------------------------Motivating Example----------------------------------------------------------------------------
\section{Motivating Examples} \label{Motivating Examples}

Before delving into the proposed approach, 
please first allow us to outline a brief introduction to smart contracts and elaborate 
on two motivating examples for our study. 

\textbf{Smart Contracts.}\quad Smart contracts are runnable code that executes 
on blockchain systems. 
They encode the user-defined rules for managing assets into 
source code~\cite{liu2023rethinking}, enabling a variety of decentralized 
and safety-guaranteed applications. 

However, many smart contracts suffer from severe vulnerabilities, and 
\textit{cloned code} may propagate vulnerabilities from the 
\textit{original} smart contract function to the \textit{plagiarized} smart contract function. 
Below, we present two real-world examples of security incidents caused by code cloning in smart contract functions. 

\textbf{Code Clone in DAPP.}\quad
Fomo3D~\cite{Fomo3D} is a highly popular Ponzi scheme-style game. It amassed an 
entry capital exceeding 40,000 Ether and eventually became a phenomenal game, with 
participation peaking at over 18,000 times per day. 
Later on, many games similar to Fomo3D emerged, exhibiting plagiarism through direct 
reuse of the source code of Fomo3D. 
Such plagiarism games include FoMoJP~\cite{FomoJp}, RatScam~\cite{RatScam}, 
SuperCard~\cite{SuperCard}, FoMoGame~\cite{FoMoGame}, and Star3Dlong~\cite{Star3Dlong}. 
Unfortunately, hackers identified a vulnerability in the original airdrop mechanism of Fomo3D~\cite{Fomo3dattackedevent} and profited over $\$$3.68 
million from Fomo3D by exploiting this vulnerability. 
Almost all the awkward imitators were also exposed to those attacks due to code 
plagiarism. 
LastWinner~\cite{LastWinner}, one of the most successful imitators of Fomo3D, was 
attacked later, incurring a loss of 12,948 ETH, valued at $\$$18 
million~\cite{Lastwinnerattackedevent}. 

\begin{figure}[t]
  \centering
    \subfloat[\textbf{a. The original BEC function}\label{fig4_1}]{
      \includegraphics[width=1\linewidth]{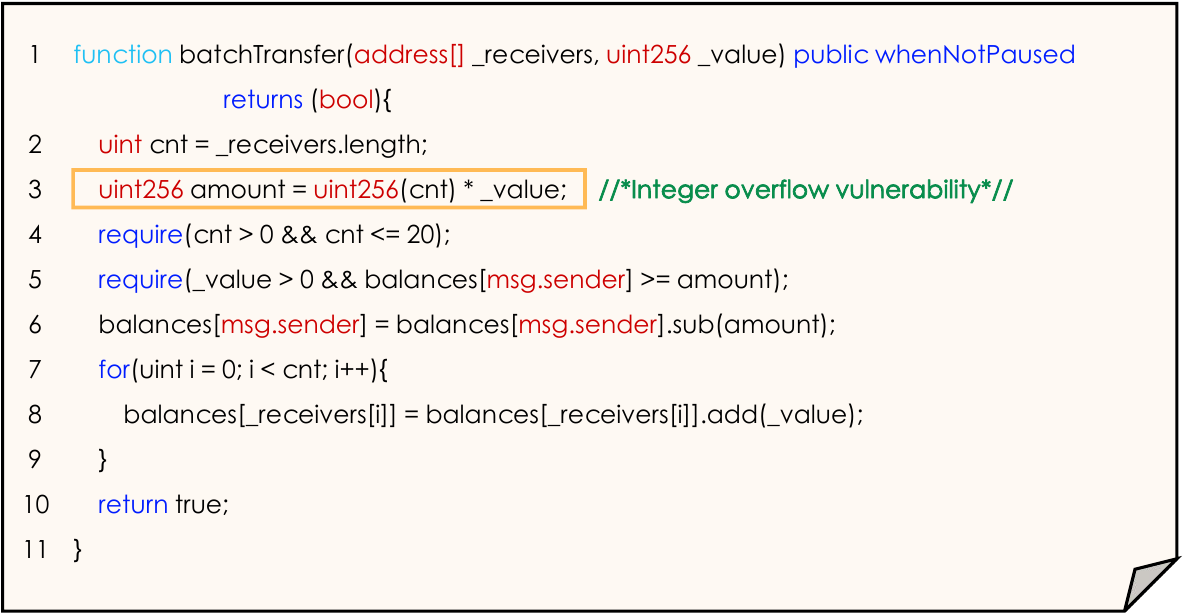}
    }\\
    \subfloat[\textbf{b. The modified function}\label{fig4_2}]{
      \includegraphics[width=1\linewidth]{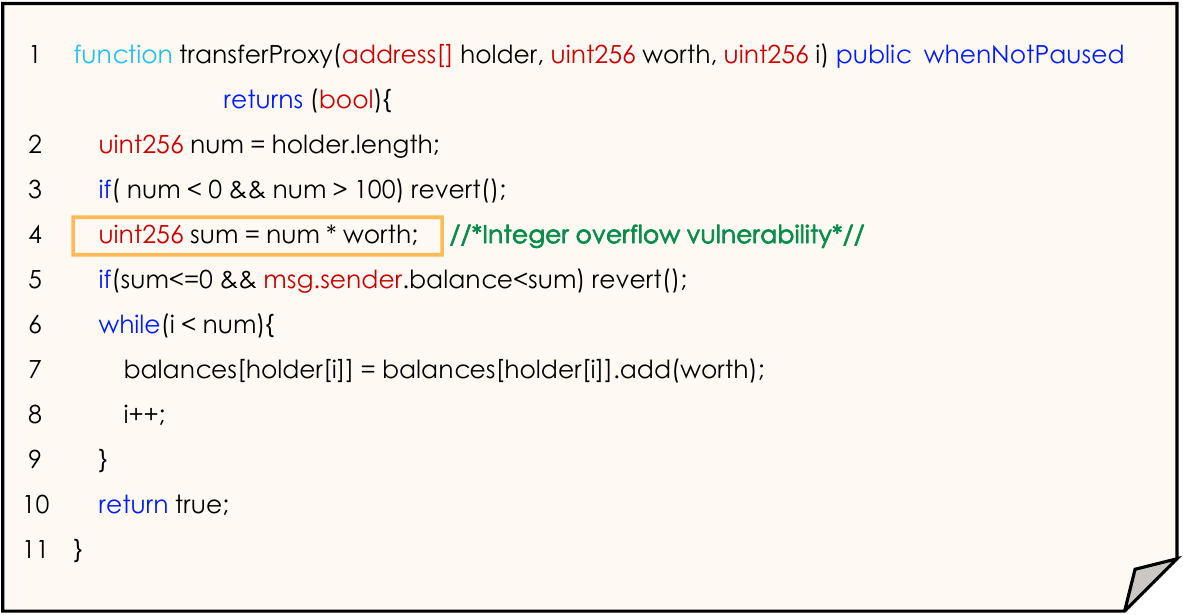}
    }
  \caption{\textbf{The original BEC transfer function and a plagiarized function from it.}}
  \label{twofunctions}
\end{figure}

\textbf{Code Clone in DeFi.}\quad
BEC (Beautiful Ecological Chain)~\cite{Beautychain}, which aims to establish 
``The world's first beautiful ecological chain platform based on blockchain 
technology,'' gained significant traction in the beauty industry. 
In 2018, the issuance of $7$ billion worth of BEC tokens resulted in a market 
value exceeding $28$ billion. 
Notably, a critical \textit{integer overflow vulnerability} in the 
BEC smart contract function (as indicated in line 3 of the orange box in 
Figure~\ref{twofunctions}\textbf{.a}) was exploited, resulting in a significant theft of tokens 
and a drastic market value devaluation of BEC. 
Similar vulnerabilities subsequently afflicted platforms like 
SMT~\cite{lee2021blockchain}, UGT~\cite{UGT}, and FNT~\cite{FNT}, all stemming from 
the same vulnerability exploit. 
In fact, the code of UGT is a direct copy of the code of SMT, and the transfer 
functions in all those platforms are remarkably similar.

% \begin{figure}[t]
%   \centering
%     \begin{subfigure}{\linewidth}
%       \includegraphics[width=1\linewidth]{figure/fig4-1-code.pdf}
%       \caption*{\textbf{a. The original BEC function}}
%       \label{fig4_1}
%     \end{subfigure}
%     \begin{subfigure}{\linewidth}
%       \includegraphics[width=1\linewidth]{figure/fig4-2-code.pdf}
%       \caption*{\textbf{b. The modified function}}
%       \label{fig4_2}
%     \end{subfigure}
%   \caption{\textbf{The original BEC transfer function and a plagiarized function from it.}}
%   \label{twofunctions}
% \end{figure}

% \begin{figure*}[t]
%   \centering
%   \setlength{\abovecaptionskip}{0.4cm} 
%     \begin{subfigure}{\linewidth}
%       \centering
%       \includegraphics[width=0.75\linewidth]{figure/ASTa2.pdf}
%       \caption*{\textbf{a. The AST of the original BEC function}}
%       \label{fig4_1}
%     \end{subfigure}
%     \begin{subfigure}{\linewidth}
%       \centering
%       \includegraphics[width=0.75\linewidth]{figure/ASTb2.pdf}
%       \caption*{\textbf{b. The AST of the modified function}}
%       \label{fig4_2}
%     \end{subfigure}
%   \caption{\textbf{The original BEC transfer function and a plagiarized function of it.}}
%   \label{example}
% \label{twoASTs}
% \end{figure*} 

\begin{figure*}[t]
  \centering
  \setlength{\abovecaptionskip}{0.4cm} 
    \subfloat[\textbf{a. The AST of the original BEC function}\label{fig4_1}]{
      \includegraphics[width=0.75\linewidth]{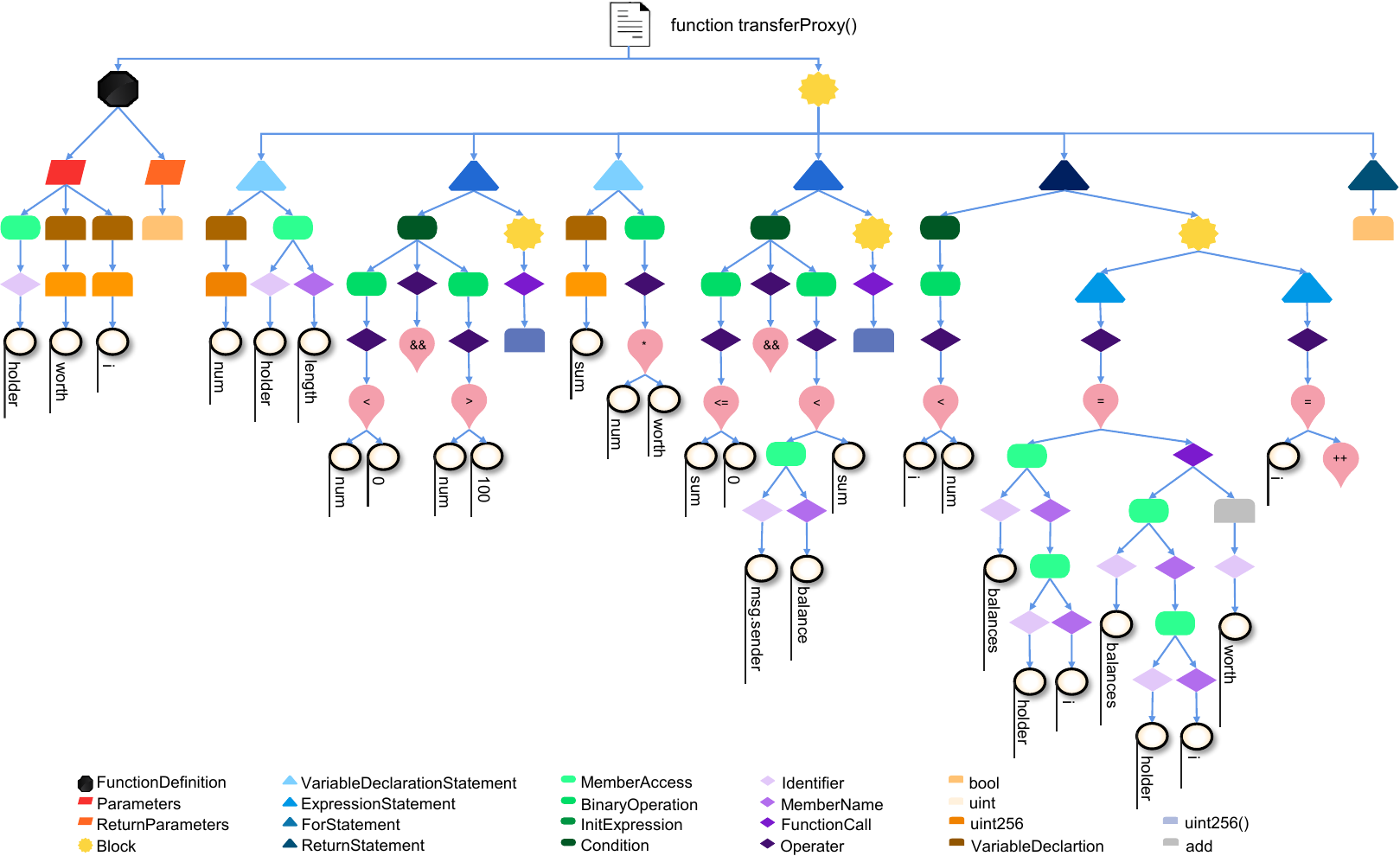}
    }\\
    \subfloat[\textbf{b. The AST of the modified function}\label{fig4_2}]{
      \includegraphics[width=0.75\linewidth]{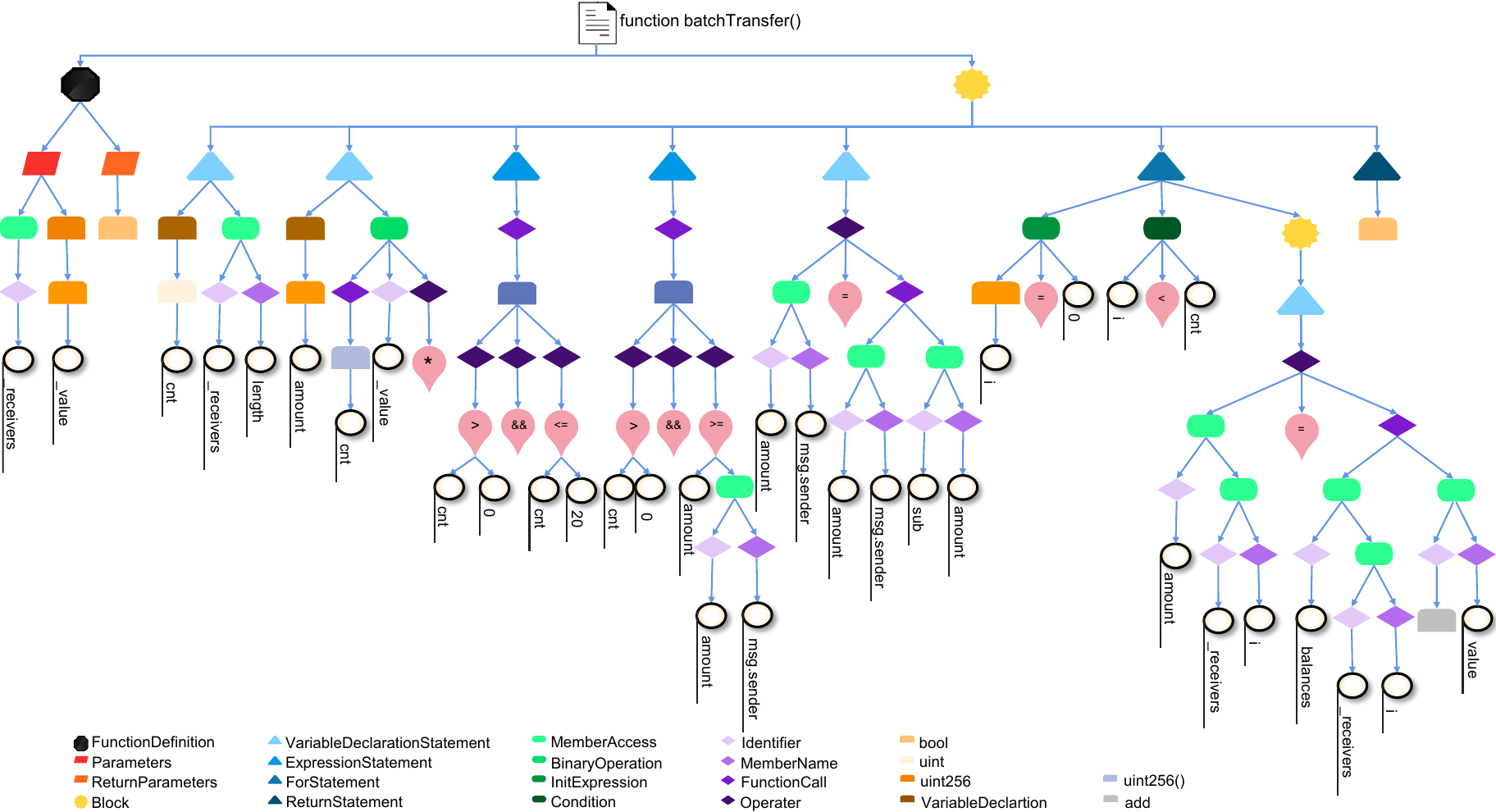}
    }
  \caption{\textbf{The original BEC transfer function and a plagiarized function of it.}}
  \label{example}
  \label{twoASTs}
\end{figure*}

\textbf{Challenges.}\quad 
Interestingly, we observe that to precisely identify code clones, such as 
the clone between the original BEC function, 
namely the \textit{batchTransfer} function, 
in Figure~\ref{twofunctions}\textbf{.a} and the modified function, 
namely the \textit{transferProxy} function, in Figure~\ref{twofunctions}\textbf{.b}, 
two critical factors must be considered: 
\textbf{(1)} 
The AST (abstract syntax tree) of a smart contract function is often large 
and complex, posing inherent challenges for similarity analysis. 
To make things worse, smart contract code of similar semantics might have 
evidently different AST structures. 
For example, Figure~\ref{twoASTs}\textbf{.a} and Figure~\ref{twoASTs}\textbf{.b} illustrate the 
ASTs of the original BEC function and the modified function, respectively. 
We can observe that the ASTs of the two functions are quite large even though 
they have only eleven lines of code. 
The semantics of these two functions are very similar, 
but their AST structures are significantly different. 
\textbf{(2)} Most existing methods adopt random forests or deep neural 
networks as the classifier to determine whether two smart contract functions 
are similar. 
The performance of these classifiers heavily relies on the setting 
of model hyperparameters. 
However, the hyperparameter space is extensive, making it challenging to 
find optimal hyperparameters. 

These motivate us to come up with two novel designs: 
\textbf{(1)} For similarity detection, we propose to break down the AST 
(abstract syntax tree) of a smart contract function into smaller and more 
elementary units, \textit{i.e.}, statement trees. 
Thereafter, we compute similarities between each statement tree from a function $A$ 
and each statement tree of the other function $B$. 
By explicitly computing the similarity between each pair of statement 
trees across the two functions, we are able to explain the reasons behind 
the similarity scores at the statement level, achieving interpretable 
similarity detection. 
\textbf{(2)} 
We propose a hyperparameter evaluation neural network and a diffusion probabilistic 
model~\cite{ho2020denoising} to search for the optimal hyperparameters of the 
classifier. 
Theoretically, we derivate the hyperparameter sampling chain process. 
We also devised a method to extract seven types of category features based on statement trees 
and assign weights to each of them for similarity detection. 

%---------------------------------------Method Overview----------------------------------------------------------------------------
\section{Method Overview}  \label{Method Overview}

Up to now, we have introduced two examples to help understand the problem, and explained the motivation of this paper. 
In this section, we formulate the problem and present the overall workflow of our approach \textsc{SmartDetector}. 

\subsection{Problem Formulation} 

\textbf{Problem Definition.}\quad Given the source code of two smart contract 
functions, 
we are interested in developing a fully automated approach that can 
\textit{explainably} predict whether the two functions are similar. 
Put differently, we seek to estimate the label $\hat{y}$ for a pair of functions, 
where $\hat{y} = 1$ represents the two functions are similar to each other 
while $\hat{y} = 0$ denotes they are dissimilar. 
Aside from providing the accurate similarity label $\hat{y}$, 
we also require the approach to be capable of locating the lines of similar code, 
thus providing interpretability. 

Broadly, smart-contract functions similarity detection methods can be divided into \textit{hand-crafted code patterns} based approaches and \textit{abstract syntax tree} (AST) based approaches. 
Recently, AST-based approaches achieved state-of-the-art results since AST contains richer semantic information than hand-crafted code patterns.

However, we observe that AST-based approaches still have two issues to be addressed. 
\textit{First}, current methods typically convert the two given functions into two ASTs and 
directly compare the two trees. 
However, the entire AST of a function is usually large and complex, 
as exemplified by the two trees in Figure~\ref{twoASTs}\textbf{.a} and Figure~\ref{twoASTs}\textbf{.b}. 
It might be overly ambitious to expect the model to directly infer the similarity 
between two large ASTs, as it involves handling numerous elements and considering 
subtle logical details. 
\textit{Second}, 
many current approaches utilize random forests or deep neural networks as 
classifiers to ascertain the similarity between two smart contract functions. 
The effectiveness of these classifiers largely hinges on the configuration of 
model hyperparameters. 
Finding optimal hyperparameters is challenging due to the extensive 
hyperparameter space. 

\begin{figure*} [ht]
  \centering 
  \setlength{\abovecaptionskip}{0.4cm}
  \includegraphics[width=18cm]{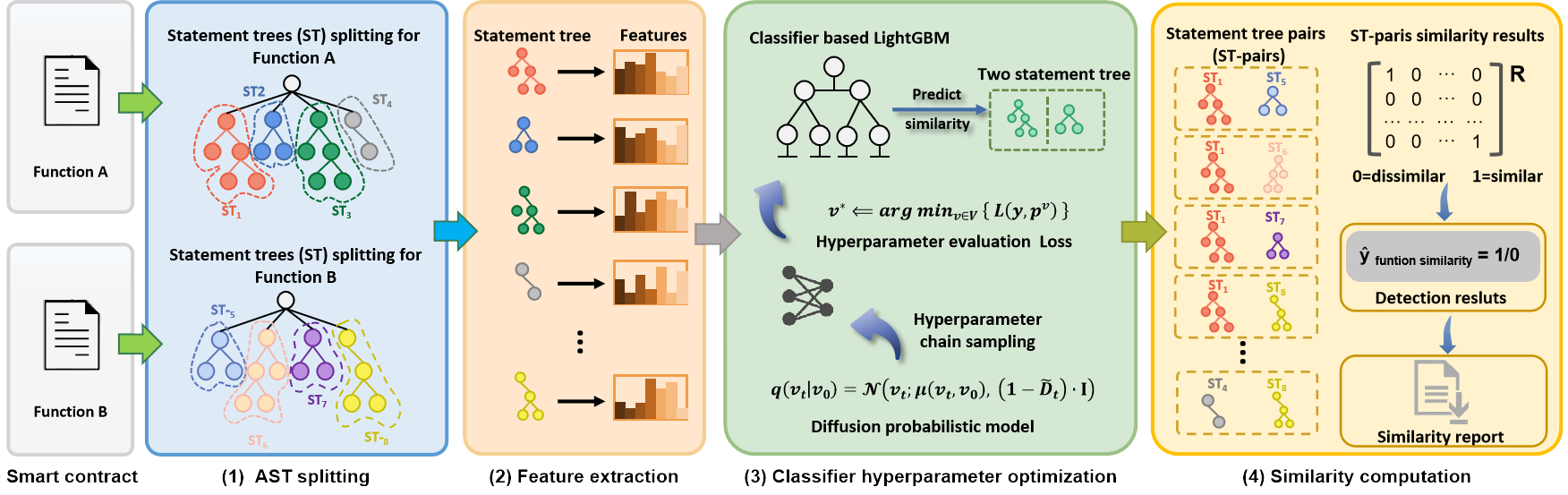} 
  \caption{\textbf{The workflow of SmartDetector in identifying the similarity between two smart contract functions.}}
  \label{workflow} 
\end{figure*}

\subsection{Workflow of SmartDetector}  \label{OverviewofInDetector}
\textbf{Method Overview.}\quad At a high level, our method 
comprises four phases: 
\textbf{(1)} In the \textit{AST splitting} phase, 
given two smart contract functions $A$ and $B$, \textsc{SmartDetector} decomposes the 
two large ASTs of the two functions into two sets of smaller (statement-level) trees, 
\textit{termed as statement trees}. 
\textbf{(2)} In the \textit{statement tree feature extraction} phase, 
we empirically investigate more than forty node types in the 
statement tree and propose to classify the nodes into seven categories. 
We develop a \textit{feature exaction tool} to extract the features for 
a given statement tree. 
Specifically, we extract the node feature for each node in the statement tree, and concatenate the node features of the 
\textit{nodes from the same category} into a category-level feature. 
The seven category-level features collectively form the feature set of 
the statement tree. 
\textbf{(3)} In the \textit{classifier hyperparameter optimization} 
phase, we train a classifier that absorbs two statement trees as inputs and outputs a probability indicating whether the two trees are similar. 
We propose a hyperparameter evaluation neural network and a 
diffusion probability model to predict the optimal hyperparameters for our 
classifier. 
Theoretically, we derivate the hyperparameter sampling chain process 
to search for the optimal classifier hyperparameters.  
\textbf{(4)} In the \textit{similarity computation} phase, 
for two sets of statement trees corresponding to the two functions $A$ and $B$, 
we compute the similarity between each statement tree of $A$ and each 
statement tree of $B$ using the classifier. 
\textsc{SmartDetector} aggregates the similarity results of all 
statement tree pairs to predict whether the two functions are similar. 
As a result, \textsc{SmartDetector} generates a similarity detection report that provides 
the similarity result for the two functions and highlights the line number(s) of similar code. 
We would like to highlight that our approach has an edge in explicitly 
addressing the common interpretability issue of smart contract 
similarity detection, as well as mathematically deducing the sampling 
chain process in classifier hyperparameters selection to optimize the classifier. 

We take the two smart contract functions of Figure~\ref{twofunctions} as an 
example to show the workflow of \textsc{SmartDetector}. 
Technically, as shown in Figure~\ref{workflow}, 
\textbf{Step 1:} given two smart contract functions $A$ and $B$ 
with their source code as inputs, 
\textsc{SmartDetector} first transforms the two functions into two ASTs, 
then decomposes the two large ASTs into 
two sets of statement trees (ST): $T_{A}=(ST_{1}, ST_{2}, \ldots, ST_{8}), T_{B}=(ST_{9}, ST_{10}, \ldots, ST_{17})$ 
by post traversal. 
\textbf{Step 2:}  
\textsc{SmartDetector} utilizes our developed \textit{feature extraction tool} to extract node features for each statement tree. 
\textbf{Step 3:} 
\textsc{SmartDetector} trains a classifier termed \textit{Smart-lightGBM} that can predict whether two statement trees are similar. 
\textsc{SmartDetector} optimizes the hyperparameters of \textit{Smart-lightGBM} using diffusion probabilistic model. 
\textbf{Step 4:} 
\textsc{SmartDetector} computes the similarity of each statement tree pair across the two given functions $A$ and $B$ 
utilizing \textit{smart-lightGBM}.
\textsc{SmartDetector} then stores the similarity prediction results of each statement tree pair 
in a matrix $R$ and aggregates the elements in $R$ to obtain the final similarity result. 
\textsc{SmartDetector} provides a similarity detection report, indicating the similarity result of the two functions and pointing out the specific similar code lines.

%-----------------------------------------------Design of statement tree-----------------------------------------------
\section{Design of statement tree}\label{Design of statement tree}

In this section, we outline the design philosophy behind the statement tree and 
elucidate the six types of statement trees we have crafted.

\textbf{Design philosophy of statement tree.}  
The abstract syntax tree (AST) of a smart contract function is often intricate and large, posing a challenge for models to detect similarity directly through comparing the two ASTs. 
To address this, we partition the large and complex AST into smaller elementary units. 
It is worth noting that determining the granularity of AST splitting is a non-trivial task. 
In this work, we opt for statement-level trees, which aligns with the fact that statements are the basic blocks that constitute the semantics of function execution. 
Specifically, we observe that a statement serves as the fundamental execution unit of a function. 
Drawing inspiration from this observation, we divide a large AST into smaller trees, referred to as \textit{statement trees}, at the granularity of individual statements. 
\textcolor{black}{This design enables more targeted and semantically meaningful comparisons, reducing the noise introduced by structurally irrelevant components in the full AST.
We also experimented with other granularities, including node-level, code-block-level, and the full AST-level. 
}
Our results demonstrate that statement-level yields the most significant performance improvements (detailed experiments are presented in Section~\ref{AblationStudy}). 

\textbf{Statement tree definition.} 
More specifically, \textsc{SmartDetector} decomposes the AST of a function into multiple statement trees, each statement tree corresponding to a line of code or a loop block in the function. 
We empirically scrutinized over 22,000 
real-world smart contracts and designed six types of statement trees 
according to the operation types of the corresponding code. 
Below, we present the explanations and the details of the six types 
of statement trees. 

\textbf{\textsl{Statement Tree Type 1: Variable Definition.}} 
The statement tree of \textit{Variable Definition} is 
designed to represent a variable definition. 
For instance, statement \textsf{`int x;'} is represented as a Variable 
Definition statement tree. 
Variables in smart contracts encompass a range of 
data types, including \textit{integer}, \textit{strings}, and 
\textit{Boolean}, to name a few. 

\textbf{\textsl{Statement Tree Type 2: Assignment Operation.}} The 
statement tree of \textit{Assignment Operation} is 
tailored for denoting an assignment operation. 
For instance, statement \textsf{`x = x + 1;'} is modeled as an 
Assignment Operation statement tree. 

\textbf{{\textsl{Statement Tree Type 3: Conditional Block.}}} The statement tree of \textit{Conditional Block} is used to represent a section of code that executes distinct operations based on different conditions. Typically, an \emph{if} statement and its branches are represented as a Conditional Block statement tree. 

\textbf{{\textsl{Statement Tree Type 4: Control Loop.}}} The 
statement tree of \textit{Control Loop} is utilized to represent a specific block of 
code that is executed multiple times until certain control conditions are met. 
A Control Loop statement tree encompasses loop control conditions 
and the operations to be executed {while} the conditions are satisfied. 
Typically, a \emph{while} statement or \emph{for} statement 
is represented as a Control Loop statement tree. 

\textbf{\textsl{Statement Tree Type 5: Function Call.}} The 
statement tree of \textit{Function Call} 
is designed to represent a call to a built-in function or a user-defined function. 
For instance, a statement invoking the build-in \textit{sha3} function to compute the \textit{sha-3} hash value is classified as a Function Call statement tree. 

\textbf{\textsl{Statement Tree Type 6: Other Operation.}} A statement not covered by any previous statement tree types is modeled 
as an \textit{Other Operation} statement tree.   For example,  statement \textsf{`return false;'} is modeled as an 
Other Operation statement tree. 
\textcolor{black}{This category also includes other uncommon but syntactically valid 
statements such as \textsf{`revert()'}, \textsf{`emit Event()'}, \textsf{`break'}, and \textsf{`continue'}. 
Due to their relatively simple structure and limited impact on similarity detection performance, grouping these statements under this type streamlines the model while preserving generality.}

\textcolor{black}{\textbf{Handling of compound statements and external calls.}
\textsc{SmartDetector} assigns the statement tree type based on the outermost syntactic structure. For example, an \textit{if} statement containing a function call is classified as a Conditional Block (Type 3), with the function call retained as a subtree node. Similarly, external calls like \textsf{`address.call(...)'} are treated as Function Call trees (Type 5) if standalone, or embedded within the corresponding control structure when nested. This hierarchical approach preserves semantic context without flattening complex execution flows.}

Interestingly, we empirically observe that the majority of statement trees extracted from 347,418 
smart contract functions fall into the first five statement tree types introduced above. 
Specifically, statistics reveal that these five types account for 96.3\% of all statement trees.

%---------------------------------------Hyperparameters Optimization ----------------------------------------------------------------------------
\section{Hyperparameters Optimization} \label{Hyperparameters Optimization}
Up to now, we have articulated the definitions of different statement trees. We developed a \emph{statement tree extraction tool} that automatically transforms a smart contract function into a set of statement trees. 

We advocate for performing similarity computation at the fine-grained statement-tree level so that we could precisely locate similar statements and deliver explainability. Given two statement trees, we propose a classifier, Smart-lightGBM, to evaluate their similarity. 
Smart-lightGBM outputs a probability indicating whether two statement trees are similar.

\textit{LightGBM} is widely used for classification. It adopts a gradient-boosting framework that utilizes tree-based ensemble learning algorithms. We select LightGBM for two reasons: (1) it can quickly process large datasets containing millions of instances, and (2) it can weigh the importance of different features, improving model interpretability.

However, \textit{LightGBM} comprises seven hyperparameters. 
For instance,  hyperparameter \textit{num\_leaves} specifies the number of leaves in a gradient-boosting tree, and hyperparameter \textit{max\_depth} controls the maximum depth of a gradient-boosting tree. The configuration of these hyperparameters directly impacts decision boundaries and thus greatly influences prediction accuracy.

In this section, we mathematically derive a cosine-wise diffusion process for hyperparameter chain sampling to search for the optimal hyperparameters of  {LightGBM}. We term our optimized LightGBM classifier  \textit{Smart-lightGBM}. 

In our classifier, the cross-entropy loss measures the divergence between the label and the prediction. 
Our objective is to continuously search for  hyperparameters that minimize the cross-entropy loss and enable \textit{Smart-lightGBM} to achieve peak performance. 

Formally, we denote the hyperparameter space as $V$ and represent the optimal hyperparameter point in $V$ as $v^*$. 
Note that $v^*$ comprises seven hyperparameters, which correspond to the seven hyperparameters of the \textit{Smart-LightGBM} classifier. Mathematically, we are to search for $v^*$ that satisfies: 

\begin{equation}
  {v^*}\leftarrow \arg\min_{v \in V} \{\overbrace{-\frac{1}{n} \sum_{i=1}^{n} (y_i \log p^v_i +  (1-y_i) \log (1- p^v_i))}^{\textit{cross-entropy loss}}\}, 
\end{equation}
where $n$ denotes the total number of samples, each comprising a pair of statement trees. 
The term $y_i$ stands for the ground truth label for the $i^{th}$  sample, indicating whether the two input statement trees are similar. 
In addition, the term $p^v_i$ represents the predicted similarity probability for the $i^{th}$ sample under hyperparameter setting $v$, describing the similarity probability between the two input statement trees. 

\textcolor{black}{\textbf{Inherent difficulty of classifier hyperparameter traversal.}}\quad
The number of \textit{hyperparameter points} in the hyperparameter space is infinite. Meanwhile, calculating the cross-entropy loss for a single hyperparameter point is time-consuming, as it requires computing the similarity for each pair of statement trees $x_i$ and $x_j$ in the training set (under such a hyperparameter setting). 

Motivated by this, we propose introducing a hyperparameter evaluation neural network that is capable of predicting the cross-entropy loss for a hyperparameter point without actually testing the hyperparameter point on the entire training set. 
In order to train this network, we need to sample several hyperparameter points in the hyperparameter space and compute their corresponding cross-entropy losses. 
Specifically, to compute the cross-entropy loss of a hyperparameter, we first
use this hyperparameter in the \textit{Smart-lightGBM} classifier. 
Next, we calculate the cross-entropy loss for each pair of statement trees in the training set and compute the average cross-entropy loss. 
These sampled hyperparameter points and their corresponding average cross-entropy losses serve as training samples for the hyperparameter evaluation network. 
We could randomly sample $k$ (\textit{e.g.,} $k$ = 5,000) hyperparameter points, and select the ones with the smallest cross-entropy losses as seeds. 

Given these seed hyperparameter points, we sample new hyperparameter points around them to acquire a sufficient number of potentially high-quality samples. 
The most \textit{intuitive} approach for sampling hyperparameter points around a seed is to perform grid search.  
\textcolor{black}{(1) Grid search systematically explores the parameter space based on predefined ranges and step sizes. }However, grid search fails to concentrate sampling around promising seeds while diminishing  sampling around non-seed samples. 
(2) Another classic sampling method is Bayesian optimization~\cite{BayesianOptimization}. \textcolor{black}{It models the objective function using a probabilistic surrogate (typically Gaussian Processes) and selects the next hyperparameter configuration by balancing exploration and exploitation. }Yet, Bayesian optimization~\cite{BayesianOptimization} involves establishing a surrogate function that aligns with the unknown objective function and tends to converge to a local optimum. 
\textcolor{black}{To address these limitations, we design a two-stage hyperparameter optimization process, summarized as follows.}

\textcolor{black}{\textbf{Overview of the Hyperparameter Optimization Process.
}To efficiently identify the optimal hyperparameters for our Smart-lightGBM classifier, we design a two-stage process. In the first stage, we randomly sample a set of hyperparameter candidates and evaluate their true cross-entropy losses on a validation set. These (hyperparameter, loss) pairs are used to train a hyperparameter evaluation neural network, which can quickly estimate the loss of unseen hyperparameter configurations. In the second stage, we propose a cosine-wise diffusion sampling process to generate new candidate hyperparameters by exploring the regions around the most promising seeds. This strategy enables us to focus the search on high-potential areas of the hyperparameter space, overcoming the inefficiencies of grid search and the local convergence issues of Bayesian optimization.  Finally, the best-performing hyperparameters are selected based on either the predicted or actual loss values.}

\textbf{Theoretical derivation of our cosine-wise diffusion sampling process.} 
\textcolor{black}{We propose a \textit{cosine-wise diffusion process} to sample around seed points. }
More specifically, we perform the diffusion process by chain sampling around a seed point $v$ in a cosine manner, and mathematically derive the sampling distribution of this process.

Technically, we define a state function to introduce Gaussian noise around a seed point in a cosine-wise manner. 
More specifically, given a seed point $v_{t-1}$ at time step $t-1$, we aim to generate a new hyperparameter point $v_{t}$ based on $v_{t-1}$:

\begin{equation}
  v_t = \sqrt{\mathcal{D}_t} \cdot v_{t-1} + \sqrt{g_t}  \cdot \epsilon,  
\end{equation}
where $t$ is the diffusion step, and $\mathcal{D}_t=1-g_t$ denotes the drift coefficient to perform a linear transformation on $v_t$. 
We define $g_t = cos^{2}(\frac{\pi}{2} \cdot \Delta{t}{\text{T}^{-1}})$ as the diffusion schedule. 
The amplitude of $g_t$ is influenced by the noise term $\epsilon$, whereas $T$ denotes the total number of time steps, and $\Delta{t}$ signifies the interval between the current time step and the initial time step.

Upon adding Gaussian noise at time step $t$, we derive a new distribution $\mathcal{N}$ for sampling. 
Formally, we denote the sampling probability distribution of $v_{t}$ given $v_{t-1}$ as $q (v_t \vert v_{t-1})$:

\begin{equation}
  \begin{split}
  q (v_t \vert v_{t-1}) 
  = & \mathcal{N}(v_t; \underbrace{\mu(v_t| v_{t-1})}_{\textit{mean}}, \underbrace{cos^{2}(\frac{\pi}{2} \cdot \Delta{t}{T^{-1}})\mathbf{I} }_{\textit{variance}}), \\  
\end{split}
\end{equation}

\begin{equation}
  \begin{split}
 {where} \quad \mu(v_t| v_{t-1}) = \sqrt{1- cos^{2}(\frac{\pi}{2} \cdot \Delta{t}{T^{-1}})} \cdot v_{t-1}. 
  \end{split}
\end{equation}

Here, $\mu(v_t| v_{t-1})$ represents the mean of $\mathcal{N}$ and $cos^{2}(\frac{\pi}{2} \cdot \Delta{t}{T^{-1}})\mathbf{I}$ denotes the variance.
The \textit{diffusion process} obeys a Markov chain that gradually samples new hyperparameter points around the seed points according to the variance schedule, 
which is given by: 
\begin{equation}
  q (v_{1:T}|v_{0}) \doteq \prod_{t=1}^{T} q (v_t \vert v_{t-1}). 
\end{equation}

Thus, we can sample any arbitrary timestep $t$ directly conditioned on the initial seed point $v_0$: 
\begin{equation}
  \begin{split}
  v_t 
  = & \sqrt{\mathcal{D}_t} \cdot v_{t-1} + \sqrt{g_t}  \cdot \epsilon \\
  = & \sqrt{\mathcal{D}_t} \cdot v_{t-1} + \sqrt{1 - \mathcal{D}_t} \cdot \epsilon \\
  = & \sqrt{\mathcal{D}_t}\sqrt{\mathcal{D}_{t-1}} \cdot v_{t-2} + \sqrt{\mathcal{D}_t}\sqrt{1 - \mathcal{D}_{t-1}} \cdot \epsilon + \sqrt{1 - \mathcal{D}_t} \cdot \epsilon  \\
\end{split}
\end{equation}

Due to $\epsilon \sim \mathcal{N}(0,1)$, 
\(\sqrt{\mathcal{D}_t}\sqrt{1 - \mathcal{D}_{t-1}} \cdot \epsilon + \sqrt{1 - \mathcal{D}_t} \cdot \epsilon\) can be written as:

\begin{equation}
\begin{cases}
X_1 \sim \sqrt{\mathcal{D}_t}\sqrt{1 - \mathcal{D}_{t-1}} \cdot \epsilon = \mathcal{N} (0, \mathcal{D}_t(1 - \mathcal{D}_{t-1})) \\
X_2 \sim \sqrt{1 - \mathcal{D}_t} \cdot \epsilon = \mathcal{N} (0, 1 - \mathcal{D}_t) \\
\end{cases}
\end{equation}

\begin{equation}
X_1 + X_2 = \mathcal{N} (0, 1 - \mathcal{D}_t\mathcal{D}_{t-1}) \\
\end{equation}

Thus, the equation can be further simplified as

\begin{equation}
  \begin{split}
  v_t = & \sqrt{\mathcal{D}_t}\sqrt{\mathcal{D}_{t-1}} \cdot v_{t-2} + \sqrt{1-\mathcal{D}_t\mathcal{D}_{t-1}} \cdot \epsilon \\
  \end{split}
\end{equation}

Further derivation yields:

\begin{equation}
  \begin{split}
  v_t 
  = & \prod_{i=1}^{t}\sqrt{\mathcal{D}_i} \cdot v_{0} + \sqrt{1-\prod_{i=1}^{t}\mathcal{D}_i} \cdot \epsilon  \\  
  \end{split}
\end{equation}

With\ \(\widetilde{\mathcal{D}}_t \doteq \prod_{i=1}^{t} \mathcal{D}_i\), we can write the marginal:

\begin{equation}
v_t = \sqrt{\widetilde{\mathcal{D}}_t} \cdot v_0 + \sqrt{(1 - \widetilde{\mathcal{D}}_t)} \cdot \epsilon
\end{equation}

A notable property of this diffusion process is that it admits sampling $v_{t}$ at any arbitrary timestep in a closed-form distribution expression. 
Put differently, 
the sampling distribution of $v_{t}$ at timestep $t$ given $v_0$ can be directly derived as:

\begin{equation}
  \begin{split}
  q (v_t \vert v_0) 
  = & \mathcal{N}(v_t; \mu(v_t, v_0), (1 - \widetilde{\mathcal{D}}_t)\mathbf{I} ) \\
  = & \mathcal{N}(v_t; \overbrace{\sqrt{\widetilde{\mathcal{D}}_t} \cdot v_0}^{\textit{refined mean}}, \underbrace{(1 - \widetilde{\mathcal{D}}_t) \mathbf{I}}_{\textit{refined variance}}). 
  \end{split}
\end{equation}

Up to now, we have derived the sampling distribution. 
We sample new hyperparameter points according to the derived sampling distribution to acquire a sufficient number of potentially high-quality hyperparameter points. 
We compute the true cross-entropy losses of these potentially high-quality hyperparameter points. 
These hyperparameters associated with their corresponding cross-entropy losses serves to adequately train the hyperparameter evaluation network. 

%---------------------------------------Hyperparameters Optimization and Similarity Report----------------------------------------------------------------------------
\section{Feature Extraction and Similarity Computation} 
\label{Feature Extraction and Similarity Computation}
In this section, we first introduce our design of statement tree node features aimed at identifying code clone behaviors.
Then, we detail our feature extraction tool tailored to extract features from a statement tree. 
Thereafter, we explain how to aggregate statement-tree level similarity scores into a function level similarity score and provide the details of localizing code clone lines. 

\subsection{Design of Statement Tree Node Features}  
\label{Our Categorization of Statement Tree Node Types}
Given two statement trees, the classifier \textit{Smart-LightGBM} can predict whether they are similar. 
However, we cannot directly feed the two statement trees into \textit{Smart-LightGBM}. Instead, we need to extract \textit{features} from the two statement trees and feed the features as inputs. 
To accomplish this, we first conduct an empirical study to analyze useful features that help identify plagiarism behaviors.  
Then, we classify statement tree nodes into seven categories and design node features for each category, as outlined in Table~\ref{table1}. 

More specifically, we note that statement trees encompass a wide array of over forty kinds of nodes, 
each implying distinct programming constructs or functionalities. 
Our node classification is rooted in the observation that several nodes exhibit similar semantics despite their different symbols. 
For instance, the operators "+" (\textit{i.e.}, addition) and "-" (\textit{i.e.}, subtraction) typically correspond to distinct kinds of nodes in the statement tree, yet they share similar semantics due to both representing arithmetic operations between variables.
Following extensive experimentation and comparisons, we group nodes that share similar semantics into the same category, resulting in the classification of statement tree nodes into seven types: 

\begin{enumerate}
    \item \textbf{Arithmetic operator node}: Arithmetic operations (\textit{e.g.}, +, *) between variables in the AST of the code.
    \item \textbf{Member variable node}: A member variable in the AST of the code (\textit{e.g.}, member variable \textit{timestamp} in \textit{block.timestamp}).
    \item \textbf{Value node}: A constant value (\textit{e.g.}, 1, 3.68, and `abcd').
    \item \textbf{Identifier node}: The name of a function or variable (\textit{e.g.}, \textit{batchTransfer()}, \textit{x}, and \textit{i}).
    \item \textbf{Unit node}: Time units or ether units (\textit{e.g.}, seconds, days, and ether).
    \item \textbf{Data type node}: Variable type or function return type (\textit{e.g.}, \textit{int} and  \textit{string}).
    \item \textbf{Code constructs node}: Nodes representing various code constructs in the AST, including code structures (e.g., \textit{for} loop  and \textit{BinaryOperation}) and function call parameter lists.  \textcolor{black}{Note that inline assembly blocks and low-level code (e.g., EVM opcodes), which lack detailed AST structure, are also classified under this category and treated as opaque units.  }   
\end{enumerate}
Each category represents distinct types of statement tree nodes. 
Notably, our strategy of categorizing node types within abstract syntax trees is introduced here for the first time in smart-contract similarity detection.

\renewcommand\arraystretch{1.2}
\begin{table*}[t]
\centering
\footnotesize
\setlength{\tabcolsep}{2.8mm}
\caption{\textbf{\centering The attributes represented by the seven category features in the source code.}}
\begin{tabular}{cc}
\hline
\hline
\textbf{Node Category} & \textbf{Explanation}    \\ \hline \hline                                    
\textbf{Arithmetic operator node}         & Arithmetic operations (\textit{e.g.}, +, *)  between variables in the AST of the code        \\ \hline
\textbf{Member variable node}       & A member variable in the AST of the code, (\textit{e.g.}, member variable \textit{timestamp} in \textit{block.timestamp})                        \\
\hline
\textbf{Value node}            & A constant value (\textit{e.g.}, 1, 3.68, and `abcd')                          \\ 
\hline
\textbf{Identifier node}             & The name of a function or variable (\textit{e.g.}, batchTransfer(), x, and i)                             \\
\hline
\textbf{Unit node}  &  Time units or ether units (\textit{e.g.}, seconds, days, and ether)                       \\ 
\hline
\textbf{Data type node}   & Variable type or function return type (\textit{e.g.}, int, string)      \\
\hline
\textbf{Code constructs node}            & Nodes representing code structures (e.g., \textit{for} loop, \textit{BinaryOperation}) and function call parameters. \\
\hline
\hline
\end{tabular}
\label{table1}
\end{table*}

\subsection{Feature Extraction Tool}  
\label{FeatureExtractionandWeight}
Building upon the categorization of node types, we further developed a \textit{feature extraction tool} tailored for extracting features from a statement tree. 

Specifically, the \textit{feature extraction tool} employs a post-order traversal to methodically extract node features from each node in a statement tree. 
Guided by our node category design, the \textit{feature extraction tool} subsequently concatenate the features of nodes within the same category into a category-level feature. 
Following these extraction and concatenation processes, we ultimately derive seven distinct category-level features. 
Collectively, these seven category-level features constitute the feature set of the statement tree.

\subsection{Function-level Similarity Computation and Code Clone Lines Localization} \label{SimilarityPredictionandReport}

In this section, we elaborate on the assessment of function-level similarity between two smart contracts using \textit{Smart-lightGBM}. 
In particular, we also generate a detailed report that pinpoints the line number(s) of similar code, facilitating further manual review.

\textbf{Similarity computation.} 
Existing methods for smart contract similarity detection often over-prioritize the order of statements while failing to concentrate on code structure and similar code elements. 
Plagiarists can exploit this by altering the statement order to interfere with similarity detection. 

To tackle this challenge, we transform the two \textit{smart contract functions under comparison} into ASTs and further decompose each AST into a set of statement trees. 
We then calculate similarity scores for each pair of statement trees between the two functions and aggregate these scores to derive the final function-level similarity.  
By evaluating similarity at the statement tree level, we can effectively identify similar code elements and structures even when there are changes in the order of statements. 

Formally, given two smart contract functions $A$ and $B$, we initially transform them into two abstract syntax trees (ASTs). 
Subsequently, we decompose these large ASTs into two sets of statement trees (STs).
These sets are represented as $T_{1}=\{st^A_{1},st^A_{2},...,st^A_{m}\}$ and $T_{2}=\{st^B_{1},st^B_{2},...,st^B_{n}\}$, where $m$ and $n$ denote the respective numbers of statement trees in $T_{1}$ (for function $A$) and $T_{2}$ (for function $B$).

Thereafter, our classifier \textit{Smart-lightGBM} evaluates the similarity between each pair of statement trees $(st^A_i,st^B_j)$, where \( st^A_i \in T_{1} \) and \(st^B_j \in T_{2} \). 
Put differently, \textit{Smart-lightGBM} evaluates the similarity between each pair of statement trees across the two functions. 
The resulting similarity scores are recorded in a matrix \( R \in \mathbb{R}^{m \times n} \), where \( R_{ij} \) denotes the similarity score between \( st^A_i \) in \( T_{1} \) and \( st^B_j \) in \( T_{2} \).
 
\textsc{SmartDetector} then calculates function level similarity between functions A and B. 
It is important to note that functions A and B may have different numbers of statement trees, so we define similarity scores for each function separately, which are given by:

\begin{equation}
    \begin{aligned}
    s_{A} & = \frac{\sum_{i = 1}^{m}\sum_{j = 1}^{n}R_{ij}}{m}, 
    s_{B} & = \frac{\sum_{i = 1}^{m}\sum_{j = 1}^{n}R_{ij}}{n}. 
    \end{aligned}
\end{equation} 
Here, $s_{A}$ measures the proportion of statement trees in function $A$ that are similar to those in function $B$, whereas $s_{B}$ measures the proportion of statement trees in function $B$ that are similar to those in function $A$. 

We then set a threshold $\delta$, and if either  $s_{A}$ or $s_{B}$ exceeds this threshold, we predict that functions $A$ and $B$ are code clones.  

\textcolor{black}{This dual-threshold design is motivated by the common \textit{asymmetry} observed in code reuse, where simpler functions are frequently embedded fully within more complex ones. 
By evaluating similarity separately in both directions, the method can accurately identify cases where one function is partially contained within another. This prevents the similarity score from being unfairly lowered due to unmatched code segments in the larger function.}

\textbf{Code clone lines localization.} 
Recall that each statement tree corresponds to a line of code or a loop block in the smart contract function. 
By comparing each pair of statement trees between the two functions, we can identify the code lines suspected of cloning behavior. 
Leveraging this capability, our method is able to generate a detailed similarity detection report that precisely pinpoints the line number(s) of similar code.  

The report first displays a conclusion indicating whether code cloning behavior exists between the two smart contract functions. 
Then, the report also presents a table that highlights the similar code between the two functions, along with their corresponding line numbers. 

\textcolor{black}{\textit{A case study} is presented in Figure 6, which demonstrates a simplified similarity report for the \textit{batchTransfer} and \textit{transferProxy} functions shown in Figure~\ref{twofunctions}. 
Each row in the report corresponds to a pair of similar code lines between the two functions, enabling precise localization of potentially cloned code. 
This facilitates efficient manual auditing and debugging by guiding developers to relevant code segments.
For example, given that line 3 in \textit{batchTransfer()} exhibits an integer overflow vulnerability, developers can use the report to trace and inspect the corresponding line (\textit{i.e.}, line 4) in \textit{transferProxy()} to investigate whether a similar vulnerability exists.}

\begin{figure} [h]
\centering 
\setlength{\abovecaptionskip}{0.4cm}
\includegraphics[width=9cm]{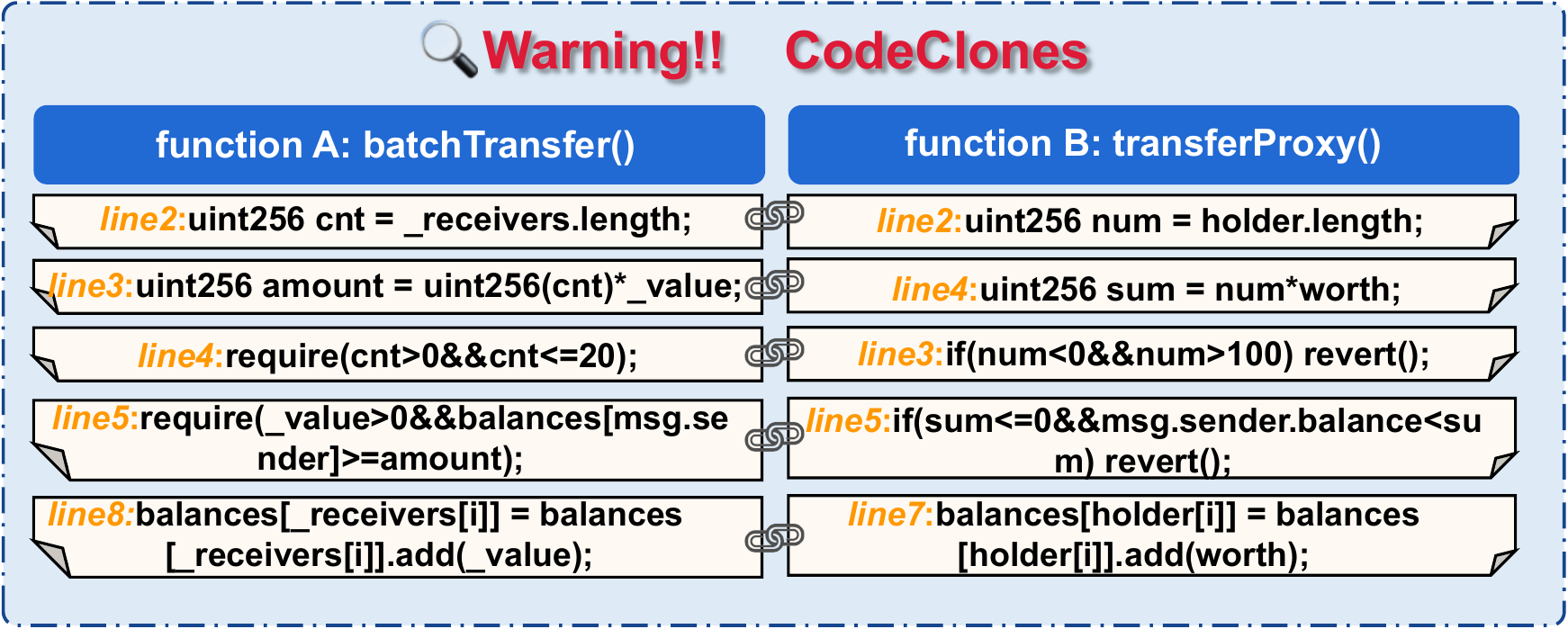} 
\caption{ \textbf{A simplified similarity detection report.} }
\label{report} 
\end{figure}

\textcolor{black}{
This code clone line localization capability ensures the interpretability of the detection results, providing clear and precise guidance for subsequent code review and security analysis.
}

%---------------------------------------Implement------------------------------------------------------------------------
\iffalse
\section{Implement} \label{Implement}

\textsc{SmartDetector} is implemented in Python and utilizes a solidity parser~\cite{solidity-parser} based on ANTLR4 grammar for extracting an AST from smart contracts. 
The code is publicly available on GitHub.
We employ a postorder traverser to break down the large AST into statement trees. 
This is particularly useful since the complex AST can strain the ability of neural networks to capture semantics. 
The traverser conducts a depth-first postorder walk in the AST, recording the information statement trees. 
We also develop an average sampling algorithm to sample different semantics and create a semantic library, assembling two statement trees into statement tree pairs and labeling whether they are cloned based on semantics.
We propose a hyperparameter evaluation neural network and a diffusion probabilistic model [10] to search for the optimal hyperparameters of our classifier \textit{Smart-lightGBM}. 
Furthermore, we develop an automated \textit{feature exaction tool} to extract and concatenate the node features into seven category-level features for each node in the statement tree. 
Notably, \textit{Smart-lightGBM} analyzes feature weights, providing insights into categorical features for similarity detection. 
\fi

%---------------------------------------Evaluate------------------------------------------------------------------------
\section{Evaluation}  \label{Evaluate}

In this section, we conduct comprehensive experiments to evaluate the proposed method. 
Overall, we seek to answer the following research questions:
\begin{itemize}
  \setlength{\itemsep}{0pt}
  \setlength{\parsep}{0pt}
  \setlength{\parskip}{0pt}
\item \textbf{RQ1:} Can \textsc{SmartDetector} effectively detect smart contract similarity at the function level? How do its precision, recall, and F1-score compare to state-of-the-art methods? 
\item \textbf{RQ2:} How efficient is \textsc{SmartDetector} in predicting smart contract code similarity compared with other methods?
\item \textbf{RQ3:} Does \textsc{SmartDetector} offer interpretability in similarity detection, and can it provide new insights?
\item \textbf{RQ4:} How much do different components of \textsc{SmartDetector} contribute to its performance in smart contract code similarity detection accuracy?
\end{itemize}
Next, we first introduce the experimental settings, followed by proceeding to answer the above research questions one by one. 
We also present a case study to allow for a better understanding of the proposed approach. 

\subsection{Experimental Datasets}\label{Experimental Datasets}

We evaluate our method on \textcolor{black}{four} benchmark datasets: \texttt{FC-pairs}, \texttt{ST-pairs}, \texttt{BL-pairs}, \textcolor{black}{and \texttt{CP-pairs}}.  
Table~\ref{dataset} presents details of the \textcolor{black}{four} datasets.

\begin{table}[h]
\centering
\caption{\textbf{Statistics of datasets for smart contract similarity detection.}}
\begin{tabular}{c|c|cc|cc}
\toprule
\multicolumn{2}{c|}{\textbf{Datasets}} & \textbf{Clone pairs} & \textbf{Non-clone pairs} & \textbf{Total} \\
\midrule
\multirow{2}{*}{\textbf{FC-pairs}} & Training set & 77,114       & 154,390         & 231,504   \\
                                   & Test set  & 34,568       & 62,196          & 96,764    \\
\midrule
\multirow{2}{*}{\textbf{ST-pairs}} & Training set & 454,824      & 745,162         & 1,199,986 \\
                                   & Test set  & 228,681      & 371,305         & 599,986   \\
\midrule
\multirow{2}{*}{\textbf{BL-pairs}} & Training set & -            & -               & -         \\
                                   & Test set  & 152          & 1,588           & 1,740     \\
\midrule
\multirow{2}{*}{\textcolor{black}{\textbf{CP-pairs}}} & \textcolor{black}{Training set} & \textcolor{black}{1,338}      & \textcolor{black}{5,354}         & \textcolor{black}{6,692}   \\
                                   & \textcolor{black}{Test set}  & \textcolor{black}{334}       & \textcolor{black}{1,339}         & \textcolor{black}{1,673}   \\
\bottomrule
\end{tabular}
\label{dataset}
\end{table}

\begin{itemize}
    \item The \texttt{FC-pairs} dataset consists of 328,268 smart contract function pairs extracted from 115,800 Ethereum smart contract functions. 
    Among them, around 111,682 pairs are labeled function clone pairs, while 216,586 are non-clone function pairs. 
    \item  The \texttt{ST-pairs} dataset contains 1,799,972 statement tree pairs extracted from 231,618 Ethereum smart contract functions. 
    Within this dataset, 683,505 pairs are labeled statement tree clone pairs and 1,116,467 are labeled non-clone pairs. 

    \item  The \texttt{BL-pairs} \cite{baselineclonepairs} dataset consists of 1,740 smart contract function pairs. 
    \textcolor{black}{The smart contract functions within \texttt{BL-pairs} dataset are collected from high-access questions on \textit{Stack Overflow}~\cite{stackoverflow}.}
    To enhance the variety of the dataset, the smart contract functions are further collected from other diverse and representative smart contract sites such as \textit{OpenZeppelin}, \textit{DASP}, and \textit{SWC-Registry}. 
    In this dataset, 152 smart contract function pairs are labeled as clone pairs, while 1,588 function pairs are labeled as non-clone pairs. 
    \item \textcolor{black}{The \texttt{CP-pairs} dataset contains 8,365 function pairs, which collected from 539 \textit{Binance Smart Chain} and 418 \textit{Polygon} smart contracts.
    Within this dataset, 1,673 pairs are labeled function clone pairs and 6,692 are labeled non-clone pairs. 
    These contracts were sourced mainly from multiple open-source repositories and verified project deployments, ensuring diversity in both contract logic and structure.}

\end{itemize}

The \texttt{BL-pairs} dataset was released in 2023 by \cite{baselineclonepairs}. 
The \texttt{FC-pairs} and \texttt{ST-pairs} datasets were collected from all open-sourced real-world smart contracts on Ethereum. 
More specifically, we first compiled a corpus containing 347,418 functions from all open-sourced smart contracts (22,094 smart contracts). 
The \texttt{FC-pairs} dataset comprises function pairs generated by randomly sampling 115,800 functions from the corpus. 
In contrast, the \texttt{ST-pairs} dataset comprises statement tree pairs extracted from the remaining 231,618 functions in the corpus. 
These statement trees are classified into six types (see section \ref{Design of statement tree} for details). 
To ensure balance, equal sampling was employed to select statement trees of the six types.

\textcolor{black}{Sixteen experienced \textit{Solidity} developers were ultimately enlisted to label and organize the data for the \texttt{FC-pairs} and \texttt{ST-pairs} datasets. 
Given the large scale of the datasets, we employed heuristic strategies to assist manual labeling and improve efficiency. 
Functions with significant functional differences or unrelated logic were categorized into distinct groups, and pairs across groups were generated as high-confidence negative samples. 
We also considered functions from different forks of the same project. Corresponding functions across these forks, such as those in ERC20 variants, tend to be highly similar. Their pairs were pre-labeled as positive samples and manually verified afterward.
Through this hybrid process combining heuristic strategies with subsequent expert manual verification, we ensured that our datasets are both large-scale and of high quality.
}

We released the datasets at
\url{https://github.com/alllgi/SmartDetecter}, hoping to facilitate future research.

\subsection{Experimental Settings}\label{Experimental Settings}
\textbf{Baselines.}  
We benchmark our proposed method against nine representative methods. 
These compared methods are state-of-the-art approaches for either smart contract similarity detection or traditional language (\textit{e.g.}, \textit{Java} and \textit{C}) similarity detection. 
Interestingly, we also involve traditional language similarity detection methods in the comparison to ensure the comprehensiveness of the evaluation. 
Specifically, these nine compared methods include four graph-based method (\textit{i.e.}, \textit{Eclone}~\cite{eclone}, \textit{FCCA}~\cite{FCCA}, \textit{DeepSim}~\cite{deepsim}, and \textit{Gemini}~\cite{gemini}), four tree-based methods (\textit{i.e.}, \textit{SmartEmbed}~\cite{smartembed}, \textit{SRCL}~\cite{SRCL}, \textit{Deckard}~\cite{deckard}, and \textit{Code2vec}~\cite{code2vec}), and one large-language-model-based method (\textit{i.e.}, \textit{ZC3}~\cite{zc3}).

For the parameters of these compared methods, we use the configurations reported as optimal in their respective papers. 
It is worth mentioning that these traditional language similarity detection methods do not natively support smart contracts. 
Therefore, we extended their functionality to handle smart contracts.

All experiments are conducted on a server equipped with an Intel Core i9 CPU at 
3.3GHz, a 2080Ti GPU, and 64GB of memory. 
Besides, to ensure the fairness of the experiment, we repeat ten times for each experiment and report the averaged results. 

\textbf{Metrics.} 
We take three widely used metrics to evaluate the performance of \textsc{SmartDetector}, including \textit{precision}, \textit{recall}, and \textit{F1-score}. 

\begin{figure}[h] 
  \centering 
  \setlength{\abovecaptionskip}{0.4cm}
  \includegraphics[width=8cm]{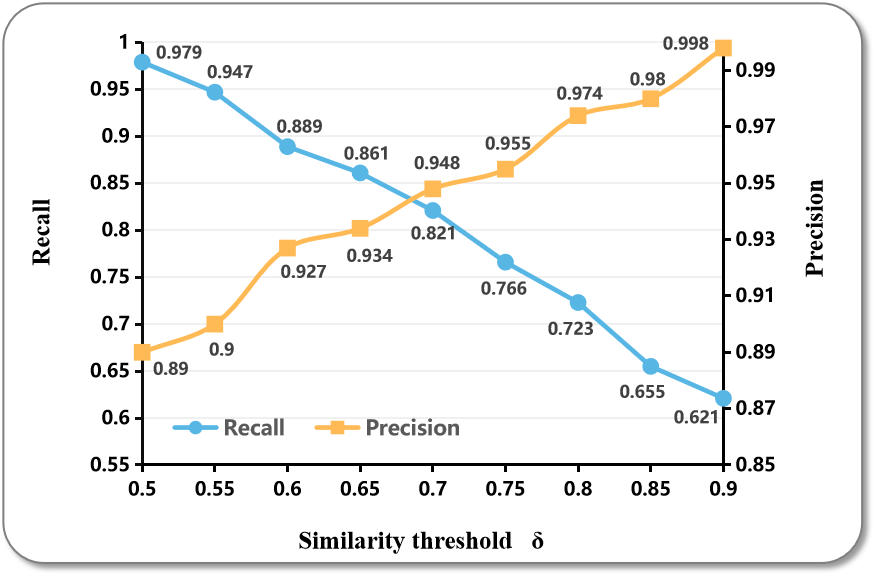} 
  \caption{ \textbf{Effect of the similarity threshold $\delta $.} }
  \label{deerta} 
\end{figure}

\textbf{Parameter Setting.}  
\textsc{SmartDetector} requires setting a similarity threshold $\delta$ to determine whether code cloning behavior exists between two functions. 
We conducted experiments to select the optimal value of the similarity threshold $\delta$ on the \texttt{ST-pairs} dataset. 
We calculated recall and precision for $\delta$ values ranging from 0.5 to 0.9 in steps of 0.05. 
The experimental results are shown in Figure~\ref{deerta}. 
It can be seen that when $\delta$ > 0.7, recall starts to decrease significantly, while precision shows only a marginal increase. 
After evaluating the trade-off between recall and precision, we ultimately choose $\delta$ = 0.7 as the default parameter.

\subsection{Overall Effectiveness (RQ1)}\label{section:6.2}
First, we benchmark \textsc{SmartDetector} against existing methods on the \texttt{FC-pairs} and \texttt{BL-pairs} datasets. 
The performance of different methods is presented in Table~\ref{performance}.

\begin{table*}[]
\centering
\caption{\textbf{Performance comparison in terms of precision, recall, and F1-score. 
A total of ten methods are investigated in the comparison, including graph-based methods, tree-based methods, large-language-model-based (\textit{i.e.} LLM-based) method, and \textsc{SmartDetector}. 
The bold and underline mark the best and second performances, respectively. }}
\begin{tabular}{|c|c|c|c|c|c|c|c|}
\hline
\multicolumn{2}{|c|}{\multirow{2}{*}{\textbf{Methods}}} & \multicolumn{3}{c|}{\textbf{FC-pairs dataset}} & \multicolumn{3}{c|}{\textbf{BL-pairs dataset}} \\ \cline{3-8} 
\multicolumn{2}{|c|}{}                                  & \textbf{Precision (\%)} & \textbf{Recall (\%)} & \textbf{F1-score (\%)} & \textbf{Precision (\%)} & \textbf{Recall (\%)} & \textbf{F1-score (\%)} \\ \hline
\multirow{4}{*}{Graph-based methods}   & Eclone~\cite{eclone}                & 63.33 & 68.63 & 65.87 & 92.11 & 61.14 & 73.49 \\ \cline{2-8} 
                               & Gemini~\cite{gemini}                & 51.57 & 57.04 & 54.17 & 85.53 & 54.39 & 66.50 \\ \cline{2-8} 
                               & DeepSim~\cite{deepsim}               & 54.84 & 43.77 & 48.69 & 89.47 & 54.84 & 68.00 \\ \cline{2-8} 
                               & FCCA~\cite{FCCA}                  & 49.96 & 56.26 & 52.92 & 84.21 & 55.65 & 67.02 \\ \hline
\multirow{4}{*}{Tree-based methods}    & SmartEmbed~\cite{smartembed}            & 59.88 & \underline{87.45} & 71.09 & \underline{96.22} & \underline{68.92} & \underline{80.31} \\ \cline{2-8} 
                               & SRCL~\cite{SRCL}                  & \underline{86.96} & 80.27 & \underline{83.48} & 94.74 & 66.36 & 78.05 \\ \cline{2-8} 
                               & Deckard~\cite{deckard}               & 60.56 & 67.57 & 63.87 & 91.45 & 58.40 & 71.28 \\ \cline{2-8} 
                               & Code2Vec~\cite{code2vec}              & 54.10 & 59.69 & 56.76 & 87.50 & 55.65 & 68.03 \\ \hline
LLM-based method    & ZC3~\cite{zc3}                   & 47.64 & 40.99 & 44.06 & 80.92 & 49.80 & 61.65 \\ \hline
\textbf{Ours}                           & \textbf{SmartDetector}         & \textbf{93.81} & \textbf{91.80} & \textbf{92.79} & \textbf{98.68} & \textbf{99.33} & \textbf{99.01} \\ \hline
\end{tabular}
\label{performance}
\end{table*}

\textbf{Comparison with graph-based methods.} 
We first compare our method with four state-of-the-art graph-based code similarity detection methods, which include: 

\begin{itemize}
    \item \textit{Eclone}~\cite{eclone}: A semantic clone detector for Ethereum smart contracts. It captures the high-level semantics of a smart contract using \textit{symbolic transaction sketches}, and combines them with other syntactic information for similarity computation to identify semantic clones. 
    \item \textit{Gemini}~\cite{gemini}: A graph-neural-network-based method for cross-platform binary code similarity detection, which represents code as a low-dimensional vector using control flow graphs to measure function similarity. 
    \item \textit{DeepSim}~\cite{deepsim}: A deep-learning-based method for measuring function similarity, which encodes code control and data flow into a semantic feature matrix to assess functional similarity. 
    \item \textit{FCCA}~\cite{FCCA}: A deep-learning-based method for code clone detection, which extracts both unstructured (\textit{e.g.}, tokens) and structured (\textit{e.g.}, control flow graph) code representations, and combines them with an attention mechanism to detect similarities. 
\end{itemize}

Quantitative results are illustrated in Table~\ref{performance}. 
From the table, we empirically observe that \textsc{SmartDetector} consistently and significantly outperforms other methods across all datasets. 
In particular, \textsc{SmartDetector} achieves 93.81\% precision, 91.8\% recall, and 92.79\% F1-score on the \texttt{FC-pairs} dataset. 
In contrast, state-of-the-art graph-based detection tools \textit{Eclone}~\cite{eclone} are 63.33\%, 68.63\%, and 65.87\%, respectively. 
Interestingly, we found that three traditional language detection methods (\textit{i.e.}, \textit{Gemini}~\cite{gemini}, \textit{DeepSim}~\cite{deepsim}, and \textit{FCCA}~\cite{FCCA}) performed worse than the smart contract detection methods \textit{Eclone}~\cite{eclone}. 
This is because traditional detection methods cannot effectively extract valuable graph structure information tailored for the smart contracts and the \textit{Solidity} language.

\textbf{Comparison with tree-based methods.} 
We further compare our method with four tree-based code similarity detection methods.
The compared methods are summarized below. 

\begin{itemize}
    \item \textit{SmartEmbed}~\cite{smartembed}: A state-of-the-art tool for detecting code clones and bugs in Ethereum smart contracts, which converts all AST elements into numerical vectors and assesses their similarities. 
    \item \textit{SRCL}~\cite{SRCL}: 
    A self-supervised learning method for smart contract representation, which extracts structural sequences from abstract syntax trees and uses discriminators to learn both local and global semantic features, enabling tasks such as code clone detection. 
    \item \textit{Deckard}~\cite{deckard}: A tree-based method for code clone detection, which represents source code as numerical vectors derived from abstract syntax trees and uses clustering to identify similar ASTs. 
    \item \textit{Code2Vec}~\cite{code2vec}: A neural-network-based method, which encodes abstract syntax tree paths into fixed-length vectors to predict semantic properties and capture code similarities.  
\end{itemize}

From the empirical evidences in Table~\ref{performance}, we observe that the proposed \textsc{SmartDetector} delivers superior performance compared to the state-of-the-art tree-based methods. 
On the \texttt{FC-pair} dataset, quantitative results reveal that \textsc{SmartDetector} achieves improvements of 21.7\%, 9.31\%, 28.92\%, and 56.76\% in F1-score over four tree-based methods (\textit{i.e.}, \textit{SmartEmbed}~\cite{smartembed}, \textit{SRCL}~\cite{SRCL}, \textit{Deckard}~\cite{deckard}, and \textit{Code2Vec}~\cite{code2vec}), respectively. 
In particular, \textit{Code2Vec}~\cite{code2vec} obtains 54.1\% precision, 59.69\% recall, and 56.76\% F1-score on the \texttt{FC-pairs} dataset, which is significantly lower than other methods. 
In consistent with the results on the \texttt{FC-pairs} dataset, \textsc{SmartDetector} keeps delivering the best performance in terms of all the three metrics on the \texttt{BL-pairs} dataset. 

Notably, we found that tree-based methods generally outperform graph-based methods. 
This is likely due to the fact that abstract syntax trees (ASTs) can more effectively represent the language structure and logical relationships inherent in the code than graph structures. 
However, most existing tree-based methods detect similar smart contracts by crudely extract code semantic information from the entire AST tree, fail to discriminate between useful and non-essential information. 
In contrast, our method decomposes the AST into a set of fine-grained statement trees. 
This strategy gives us an advantage in obtaining more precise semantic information about the code at the statement level, facilitating semantic comparison at a finer granularity. 

The empirical experimental results further show that \textsc{SmartDetector} achieves the best detection performance on all the three datasets, demonstrating that AST splitting and the optimized classifier help \textsc{SmartDetector} in precise code clone detection.

\textbf{Comparison with large-language-model-based method.} 
We also compare our methods with a large-language-model-based method \textit{ZC3}~\cite{zc3}. 
\textit{ZC3}~\cite{zc3} is a zero-shot code similarity detection method based on the large language model \textit{CodeBERT}~\cite{feng2020codebert}. 
It employs domain-aware learning and cycle consistency learning, allowing the model to efficiently detect similar smart contracts. 

We illustrate the results in terms of precision, recall, and F1-score in Table~\ref{performance}. 
These quantitative results reveal that \textit{ZC3}~\cite{zc3} does not perform well on these datasets. 
For instance, \textit{ZC3}~\cite{zc3} obtains 47.64\% precision, 40.99\% recall, and 44.06\% F1-score on the \texttt{FC-pairs} dataset. 
This may attribute to the inherent challenge that large-language-model-based methods face in effectively capturing the structural semantics and syntactic features of code, especially when compared to tree-based and graph-based methods.

\begin{figure}[h] 
  \centering 
  \color{black}
  \setlength{\abovecaptionskip}{0.4cm}
  \includegraphics[width=8cm]{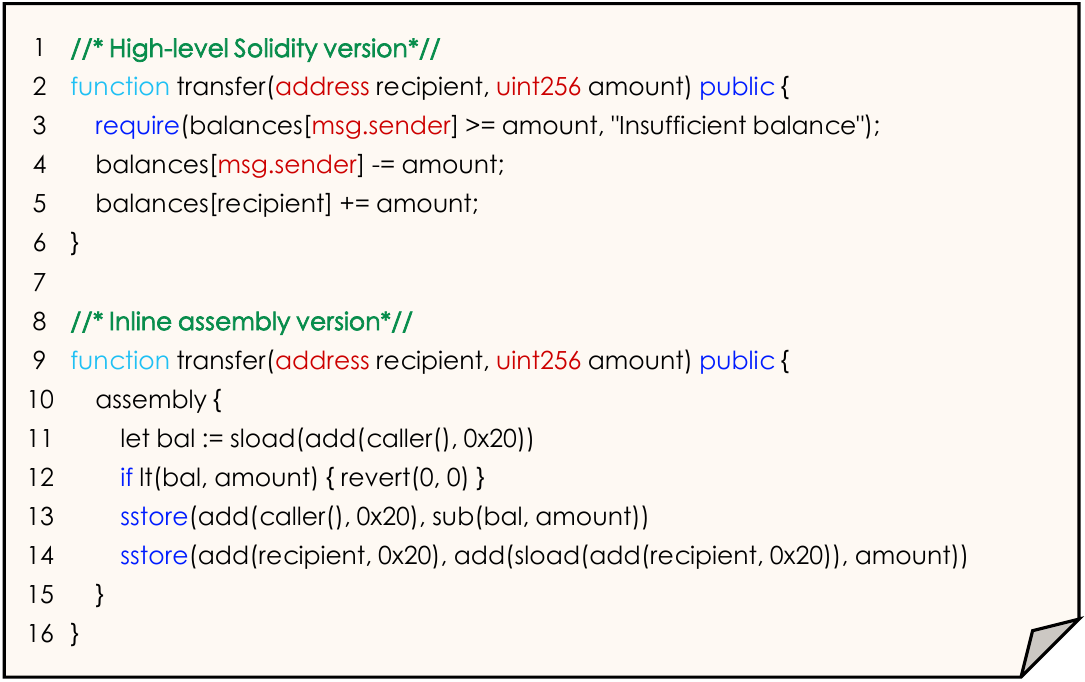} 
  \caption{ \textcolor{black}{\textbf{Two \textit{transfer} functions, one written in high-level Solidity and the other using inline assembly. The inline assembly version leads to a false negative in similarity detection due to differences in AST representations.}
} }
  \label{fn} 
\end{figure}

\textcolor{black}{\textbf{False Negative Example: Inline Assembly Obscuring Transfer Logic.}
A case of false negatives in \textsc{SmartDetector} arises from the use of inline assembly in smart contracts. Although semantically equivalent, high-level Solidity code and inline assembly differ significantly in their AST representations. For example, the two functions in Figure~\ref{fn} both implement the same transfer logic, but the inline assembly version results in a simplified AST due to its opaque nature.}

\textcolor{black}{As a result, \textsc{SmartDetector} may fail to detect the similarity, leading to a false negative. This illustrates a limitation of AST-based detection methods, where low-level assembly code and high-level code are not semantically aligned. While \textsc{SmartDetector} reduces this issue by assigning lower weights to assembly nodes, fully bridging the gap remains an open challenge and an important direction for future work.
}

\begin{figure}[h] 
  \centering 
  \color{black}
  \setlength{\abovecaptionskip}{0.4cm}
  \includegraphics[width=8cm]{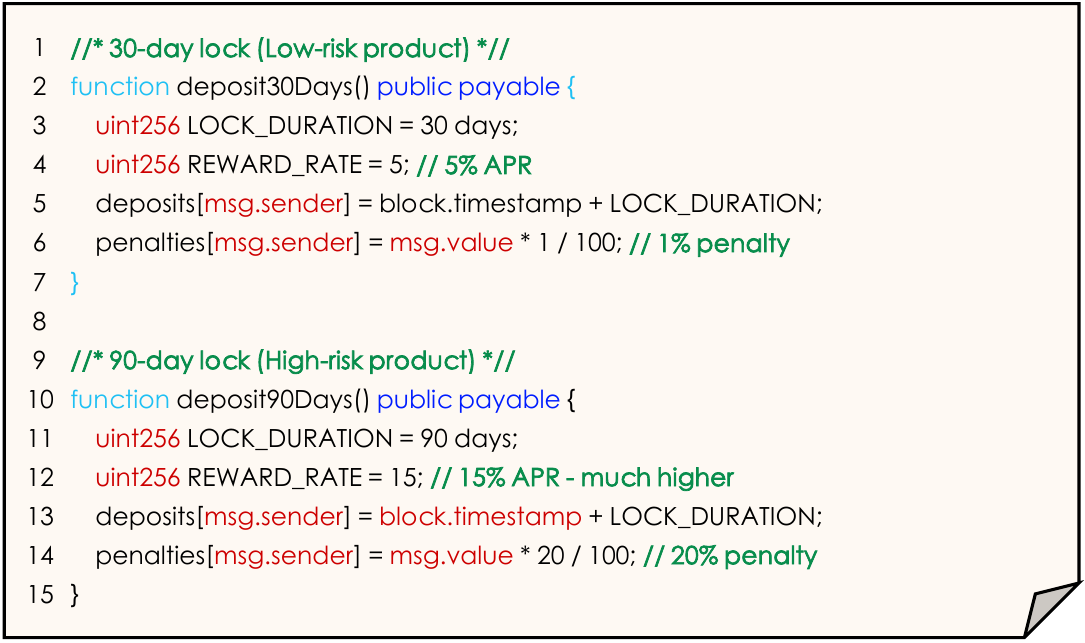} 
  \caption{ \textcolor{black}{\textbf{Comparison of two functionally similar contracts with different lock durations.}
} }
  \label{fp} 
\end{figure}

\textcolor{black}{\textbf{False Positive Example: Parameterized Differences in Lock Duration.}
Some functions share identical logic and high structural similarity, differing only in their parameter values, yet serve different business purposes. This can lead to false positives.
In Figure~\ref{fp}, the first function uses a 30-day lock with low rewards and minor penalties, while the second employs a 90-day lock with higher rewards and severe penalties, positioning it as a high-risk, high-reward product. These parameter differences lead to vastly distinct financial products, each with significantly different risk and profit profiles.
}

\textcolor{black}{ This structure is almost identical, yet the semantics differ significantly, making it prone to false positives. Bridging the gap between syntactic similarity and semantic equivalence remains an ongoing challenge.
}

\textcolor{black}{\textbf{Cross-Platform Evaluation.}
To assess the generalizability of \textsc{SmartDetector}, we extended our experiments to smart contracts deployed on the Binance Smart Chain (BSC) and Polygon. A new dataset, CP-pairs, was created, containing 8,365 function pairs sourced from 539 BSC and 418 Polygon smart contracts. 
The contracts were collected from multiple open-source repositories and verified project deployments, ensuring diversity in both contract logic and structure.}

\textcolor{black}{Experimental results show that SmartDetector maintains high performance on the CP-pairs dataset, achieving a precision of 93.8\%, recall of 90.9\%, and an F1-score of 92.1\%.
These results are consistent with those obtained on Ethereum contracts, demonstrating the robustness and generalizability of our approach across different EVM-compatible platforms.
}

\subsection{Efficiency (RQ2)}\label{Efficiency(RQ2)}
In this subsection, we systematically examine the efficiency of \textsc{SmartDetector} and compare it with other methods. 

\begin{figure*} [t]
  \centering 
  \setlength{\abovecaptionskip}{0.4cm}
  \includegraphics[width=1\linewidth]{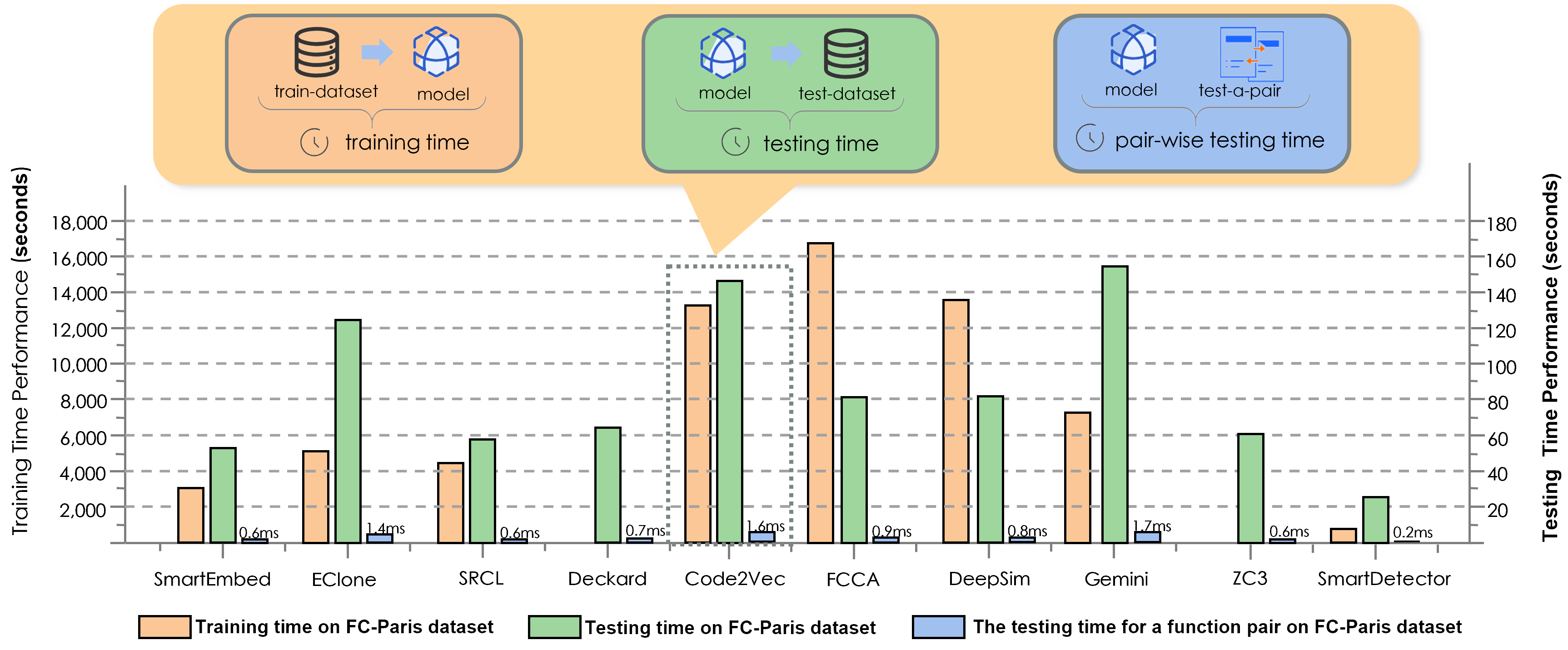} 
  \caption{ \textbf{Time performance on FC-pairs dataset.} }
  \label{time-tools} 
\end{figure*}

Specifically, we conduct experiments to measure the efficiency of \textsc{SmartDetector} by calculating the execution time on the \texttt{FC-pairs} dataset. 
To reveal the time consumption of detection methods, we examine the runtime of model training and testing, as well as the time required to test a single smart contract function pair. 
To mitigate the variation in runtime caused by the fluctuating usage status of the machine at different times, we run each method ten times and choose the average runtime as the final measure of overhead. 
Table~\ref{time} presents the comparison results of time performance for different methods. 
We further visualize the quantitative results of Table~\ref{time} in Fig.~\ref{time-tools}. 

\begin{table}[h]
  \centering
  \footnotesize
  \setlength{\tabcolsep}{3.8mm}
  \caption{\textbf{Time performance on FC-pairs dataset.}}
  \begin{tabular}{lcc}
  \hline
  \hline
  \textbf{Methods} & \textbf{Training Time} & \textbf{Testing Time}    \\ \hline
  SmartEmbed       & 2,937s±205s            & 59s±2s                      \\ \hline
  EClone           & 5,206s±364s            & 139s±9s                      \\ \hline
  SRCL             & 4,456s±234s            & 64s±3s                      \\ \hline
  Deckard          & -                      & 72s±6s                      \\ \hline
  Code2Vec         & 13,212s±489s           & 163s±92s                      \\ \hline
  FCCA             & 16,769s±3,973s         & 91s±6s                      \\ \hline
  DeepSim          & 13,545s±948s           & 86s±6s                      \\ \hline
  Gemini           & 7,312s±34s             & 172s±4s                       \\ \hline
  ZC3              & -                      & 68s±13s                       \\ \hline
  \textbf{SmartDetector}    & \textbf{842s±3s}   & \textbf{29s±1s}    \\ 
  \hline
  \hline
  \end{tabular}
  \label{time}
\end{table}

From Fig.~\ref{time-tools}, we observe that graph-based methods (\textit{i.e.}, \textit{Eclone}~\cite{eclone}, \textit{Gemini}~\cite{gemini}, \textit{DeepSim}~\cite{deepsim}, \textit{FCCA}~\cite{FCCA}) exhibit relatively higher time performance compared to tree-based methods and large-language-model-based method. 
This may due to their reliance on graph neural networks, which require more time to process extensive node and edge interactions within complex graph structures. 
As illustrated in the Table~\ref{time}, \textsc{SmartDetector} spends an average of 842 seconds during the training phase and 29 seconds during the testing phase, demonstrating significantly greater efficiency compared to other state-of-the-art methods. 
For the state-of-the-art code similarity detector \textit{SmartEmbed}~\cite{smartembed}, it takes an average of 2,937 seconds for training and 59 seconds for testing. 
In contrast, \textsc{SmartDetector} consumes approximately 3.5 times less in training and is about 2 times faster in testing.

To investigate the performance of each model in handling individual samples, we also measured the time required to test a single smart contract function pair. 
The graph-based method \textit{Gemini}~\cite{gemini} takes the longest time, averaging 1.7 ms per function pair.  
In comparison, the state-of-the-art method \textit{SmartEmbed}~\cite{smartembed} averages 0.6 ms in per function pair. 
\textsc{SmartDetector} achieves an average detection speed of 0.2 ms per function pair, significantly outperforming other compared methods.

\subsection{Interpretability Evaluation (RQ3)} 
\label{section:6.4} 
We evaluate the interpretability of \textsc{SmartDetector} from two angles. 
1) We divide an AST into statement trees, we explain how statement trees align with corresponding lines in code. 
2) We show the weights of the seven category-level features (see sections \ref{Our Categorization of Statement Tree Node Types} and \ref{FeatureExtractionandWeight} for details) in the Smart-lightGBM classifier. 

\textbf{Interpretability at the statement Level.} 
The interpretability of \textsc{SmartDetector} is rooted in our abstract syntax tree (\textit{i.e.}, AST) division strategy. 
This strategy decomposes the AST into a set of fine-grained statement trees, with each tree corresponding to a line of code or a loop block within the smart contract function. 
By calculating the statement-tree-level similarity scores between two compared smart contracts, we can identify code lines that may exhibit cloning behavior. 
This enables our method to generate a detailed similarity detection report that precisely pinpoints the line number(s) of similar code. 
Based on the accurate tracing of the detected similar code lines, we can further support manual auditing, thus ensuring the interpretability of our detection results.

\textbf{Interpretability on feature weight.}
As presented in Section~\ref{Feature Extraction and Similarity Computation}, \textsc{SmartDetector} extracts seven category-level features for statement tree nodes.  
To discern the most influential category features in code similarity detection, we tracked how frequently these features were used during training, which reflects their importance in code similarity detection. 
Table~\ref{feature-weight} lists their weights. 
We can observe that the \textit{data type node feature}, \textit{arithmetic operator node feature}, \textit{member variable node feature}, and \textit{value node feature} get higher weights, while the \textit{identifier node feature}, \textit{unit node feature}, and \textit{code constructs node feature} obtain lower weights.

\renewcommand\arraystretch{1.2}
\begin{table}[h]
\centering
\footnotesize
\setlength{\tabcolsep}{4.5mm}
\caption{\textbf{Weights of the seven category-level features in detecting similar code.}}
\begin{tabular}{ccc}
\toprule
\textbf{Rank} & \textbf{Feature} & \textbf{Weight} \\
\midrule
1 & Data type node feature            & 0.41664 \\
2 & Arithmetic operator node feature & 0.19156 \\
3 & Member variable node feature     & 0.12574 \\
4 & Value node feature               & 0.12108 \\
5 & Identifier node feature          & 0.06333 \\
6 & Unit node feature                & 0.05588 \\
7 & Code constructs node feature     & 0.02515 \\
\bottomrule
\end{tabular}
\label{feature-weight}
\end{table}

\subsection{Ablation Study (RQ4)}\label{AblationStudy}
By default, \textsc{SmartDetector} employs the \textit{proposed AST splitting strategy} to decompose the entire AST into statement trees, aiming to effectively capture both code features and structural-level syntactical information. 
It is interesting to see the effect of removing this strategy. 
Moreover, \textsc{SmartDetector} adopts a \textit{cosine-wise diffusion sampling process} to efficiently probe the optimal hyperparameters of our classifier. 
We are curious about how much this module contributes to the performance of \textsc{SmartDetector}. 
Finally, \textsc{SmartDetector} introduces a \textit{novel feature extraction mechanism} that collects features from seven specifically designed node categories, intended to enhance the extraction of crucial features within statement trees. 
It is useful to evaluate the contributions of this mechanism by removing it from \textsc{SmartDetector} as well. 
In what follows, we conduct experiments to study the three components, respectively.

\textbf{Study on the AST splitting strategy.}
\textcolor{black}{
To demonstrate how AST decomposition contributes to better detection performance, we replaced the AST splitting strategy in \textsc{SmartDetector} with three alternative methods:} (1) using the original whole AST as inputs, which is termed as \textit{SmartDetector-AF}, (2) treating each AST node as an individual input, which is termed as \textit{SmartDetector-AN}, and (3) dividing the AST into segments based on code blocks, which is termed as \textit{SmartDetector-AB}. 
Quantitative results are summarized in Table~\ref{ablation}. 
The table shows that \textsc{SmartDetector} significantly outperforms the three variants. 
This suggests the effectiveness of our AST splitting strategy.

\renewcommand\arraystretch{1.2}
\begin{table}[t]
\color{black}
\centering
\footnotesize
\setlength{\tabcolsep}{3.8mm}
\caption{\textbf{Precision, recall, and F1-score comparison (\%) between SmartDetector and its variants.}}
\begin{tabular}{lccc}
\toprule
\textbf{Methods} & \textbf{Precision} & \textbf{Recall}  & \textbf{F1-score} \\
\midrule
% Group 1: Ablation variants
SmartDetector-AF  & 91.24\%   & 86.34\%   & 88.72\%  \\
SmartDetector-AN  & 91.35\%   & 86.61\%   & 88.92\%  \\
SmartDetector-AB  & 91.45\%   & 87.24\%   & 89.30\%  \\
\midrule
% Group 2: Hyperparameter tuning variants
SmartDetector-HPO & 82.16\%   & 63.26\%   & 71.48\%  \\
SmartDetector-BO  & 90.65\%   & 86.38\%   & 88.46\%  \\
SmartDetector-GS  & 88.65\%   & 84.91\%   & 86.74\%  \\
\midrule
% Group 3: Feature ablation
SmartDetector-WFE & 87.26\%   & 47.17\%   & 61.24\%  \\
\midrule
% Group 4: Final model
\textbf{SmartDetector} & \textbf{93.81\%} & \textbf{91.80\%} & \textbf{92.79\%} \\
\bottomrule
\end{tabular}
\label{ablation}
\end{table}

\textbf{Study on the hyperparameters optimization.}
To evaluate the contribution of our cosine-wise diffusion sampling process, we first removed it and crafted \textit{SmartDetector-HPO}, which adopts the default LightGBM hyperparameter configuration. 
We further developed \textit{SmartDetector-BO} and \textit{SmartDetector-GS}, which use Bayesian optimization~\cite{BayesianOptimization} and grid search respectively for sampling hyperparameter points. 
We conduct experiments to compare our default method with these three variants to assess the impact of different hyperparameter optimization strategies.
Empirical results are shown in Table~\ref{ablation}, which reveal that \textsc{SmartDetector} outperforms \textit{SmartDetector-HPO}, \textit{SmartDetector-BO}, and \textit{SmartDetector-GS} by 21.31\%, 4.33\%, and 6.05\% in F1-score, respectively. 

\textcolor{black}{The observed performance gains may stem from the theoretical advantages of our diffusion-based optimization. Specifically, grid search incurs high computational cost and lacks adaptability, particularly in high-dimensional spaces. Bayesian optimization~\cite{BayesianOptimization}, while more efficient, depends on surrogate modeling and may struggle with scalability and convergence to global optima.
In contrast, our cosine-wise diffusion process defines a probabilistic sampling chain that directly minimizes cross-entropy loss, enabling adaptive and efficient exploration of the hyperparameter space.
These theoretical insights, supported by empirical results, suggest that our hyperparameter optimization module contributes to the improved performance of \textsc{SmartDetector}.}

\textbf{Study on category-level node features extraction mechanism.}
We further investigate the impact of the category-level node features extraction mechanism in \textsc{SmartDetector}. 
Specifically, we remove this mechanism and rely solely on statement trees as inputs. 
We refer to this new method as \textit{SmartDetector-WFE}. 
Empirical results in Table~\ref{ablation} demonstrate that \textit{SmartDetector-WFE} experiences significant performance drops compared to \textsc{SmartDetector}. 
In detail, \textit{SmartDetector-WFE} exhibits a 6.55\% decrease in precision, a 42.41\% decrease in recall, and a 30.41\% decrease in F1-score. 
These findings underscore the critical role of the category-level features extraction module in \textsc{SmartDetector}.

\section{Discussion and Future Work} 
\label{Discussion and Future Work}
\textcolor{black}{\textsc{SmartDetector} targets smart contract similarity detection and achieves promising results on benchmark datasets. Nonetheless, key challenges and potential improvements remain in the following aspects:}

\textcolor{black}{\textbf{Robustness Against Adversarial Code.} Robustness against adversarially modified or obfuscated code is crucial for practical similarity detection. Our benchmark datasets (e.g., FC-pairs, ST-pairs) include some common transformations like dead code insertion and variable renaming but may miss more advanced adversarial cases. Future work will focus on creating specialized evaluation sets with semantic-preserving transformations to better assess and improve \textsc{SmartDetector}'s resilience.
}

\textcolor{black}{\textbf{Handling of Inline Assembly Blocks. }
The Solidity compiler offers limited AST-level support for inline assembly, often treating these blocks as opaque nodes. Accordingly, \textsc{SmartDetector} abstracts inline assembly as leaf nodes under the Code Constructs category to maintain statement tree consistency and avoid parsing low-level, unstructured code. This design helps reduce false negatives by assigning lower weights to these nodes and leveraging contextual information from surrounding code.
Nonetheless, bridging the semantic gap between inline assembly and high-level Solidity remains challenging, especially in complex cases. We acknowledge this limitation and recognize it as an important direction for future work.}

\textcolor{black}{\textbf{Leveraging Similarity for Vulnerability Identification.} 
\textsc{SmartDetector} is designed to accurately detect similar smart contracts, providing a foundation for downstream vulnerability analysis. While it does not directly identify vulnerabilities, the similarity results can guide targeted security investigations. Future work will explore integrating vulnerability mining with similarity detection to enable the discovery of new vulnerable contracts.
}

\textcolor{black}{\textbf{Scalability and Real-World Deployment.} 
Direct similarity comparison faces challenges in real-world deployment due to the large number of contracts. Addressing these scalability issues and validating practical deployment strategies remain important directions for future work.
}

%--------------------------------------Conclusion--------------------------------------------------------------
\section{Conclusion} \label{Conclusion}
In this paper, we try to address the problem of smart contract similarity detection from three directions: 
1) We propose to decomposes ASTs into fine-grained statement trees based on code execution logic, which allows for effective capture of both code features and structural syntactical information. 
2) We adopt a mathematically derived cosine-wise diffusion process for hyperparameter chain sampling, crafted to efficiently probe the optimal hyperparameters of our classifier. 
3) We classify statement tree nodes into seven categories and design seven types of node features, facilitating and enhancing the extraction of crucial features from the statement trees. 
These three elements are seamlessly integrated into a novel approach, enabling statement-level similarity computation between two smart contract functions. 
Significantly, \textsc{SmartDetector} provides a detailed report that includes not only the similarity results but also the line numbers of similar code, offering statement-level interpretability for similarity detection. 
Extensive experiments on three benchmark datasets demonstrate the efficacy of our approach. 
We also discussed the efficiency and interpretability of the proposed method, as well as the contributions of its key components. 
Our implementations and datasets are released to facilitate future research.

\bibliographystyle{IEEEtran}
\bibliography{SmartDetector}

\vspace{-10pt}

\begin{IEEEbiography}[{\includegraphics[width=1in,height=1.25in,clip,keepaspectratio]{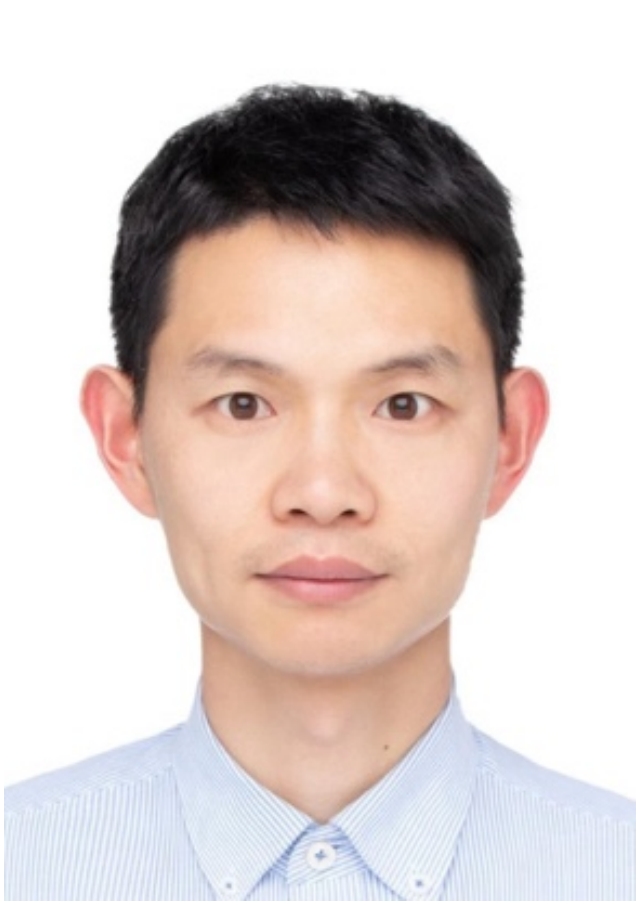}}]{Zhenguang Liu} is currently a professor of Zhejiang University. He had been a research fellow in National University of Singapore and A*STAR. (Agency for Science, Technology and Research, Singapore). He respectively received his Ph.D. and B.E. degrees from Zhejiang University and Shandong University, China. His research interests include blockchain smart contract and multimedia data analysis. Various parts of his work have been published in first-tier venues including PAMI, ACM CCS, S\&P, USENIX Security, TDSC, TIFS, CVPR, ICCV, TKDE, TIP, WWW, AAAI, ACM MM, ICLR, KDD, NeurlPS, INFOCOM, IJCAI, etc. Dr. Liu has served as technical program committee member for top-tier conferences such as CVPR, ICCV, WWW, AAAI, IJCAI, ACM MM, session chair of ICGIP, local chair of KSEM, and reviewer for IEEE PAMI, IEEE TVCG, IEEE TPDS, IEEE TIP, etc.
\end{IEEEbiography}

\vspace{-10pt}

\begin{IEEEbiography}[{\includegraphics[width=1in,height=1.25in,clip,keepaspectratio]{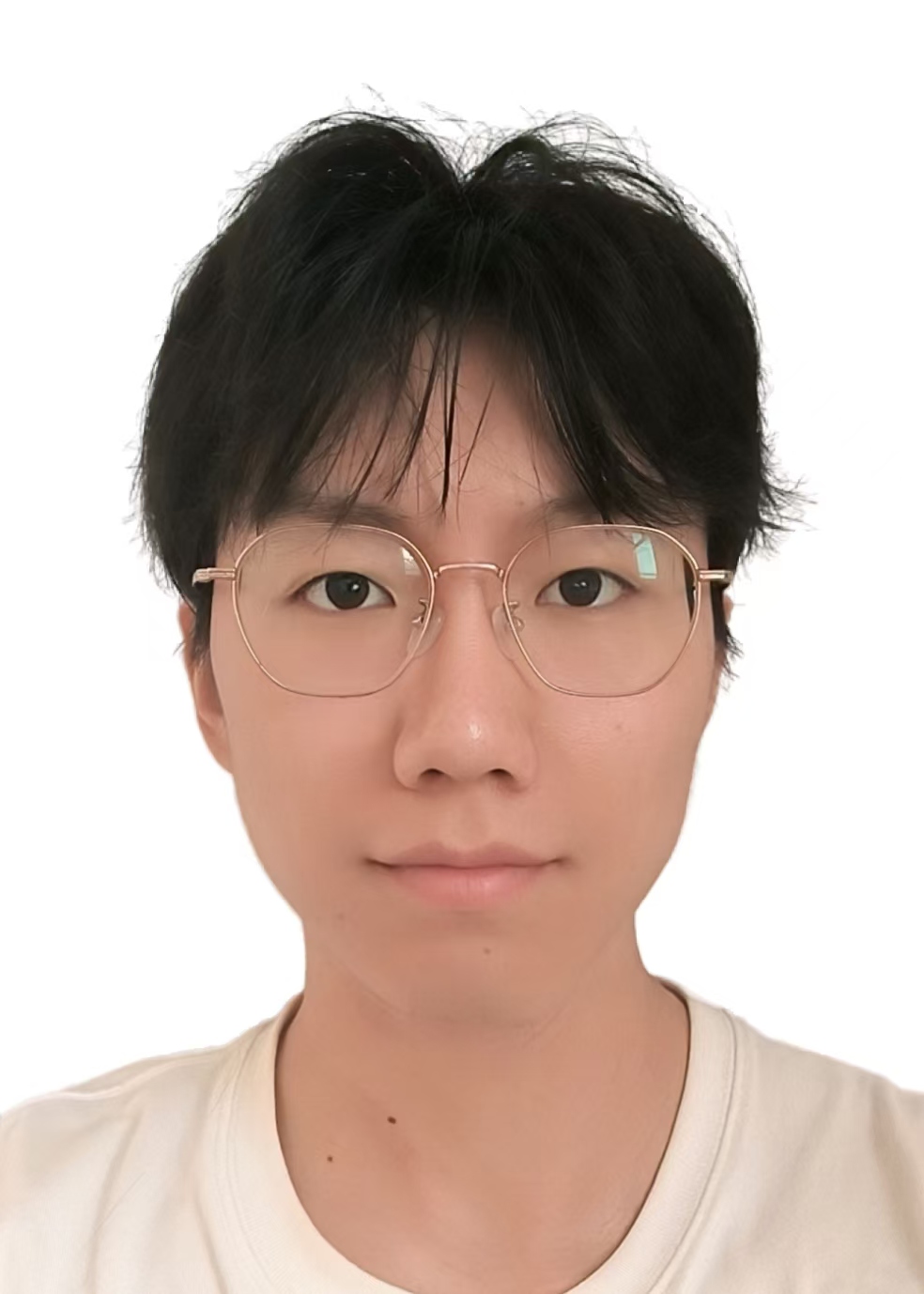}}]{Lixun Ma} 
received the B.E. degree from China University of Mining and Technology.
He is currently pursuing the Ph.D. degree with Zhejiang University.
His research interests include blockchain and smart contract security. 
\end{IEEEbiography}

\vspace{-20pt}

\begin{IEEEbiography}[{\includegraphics[width=1in,height=1.25in,clip,keepaspectratio]{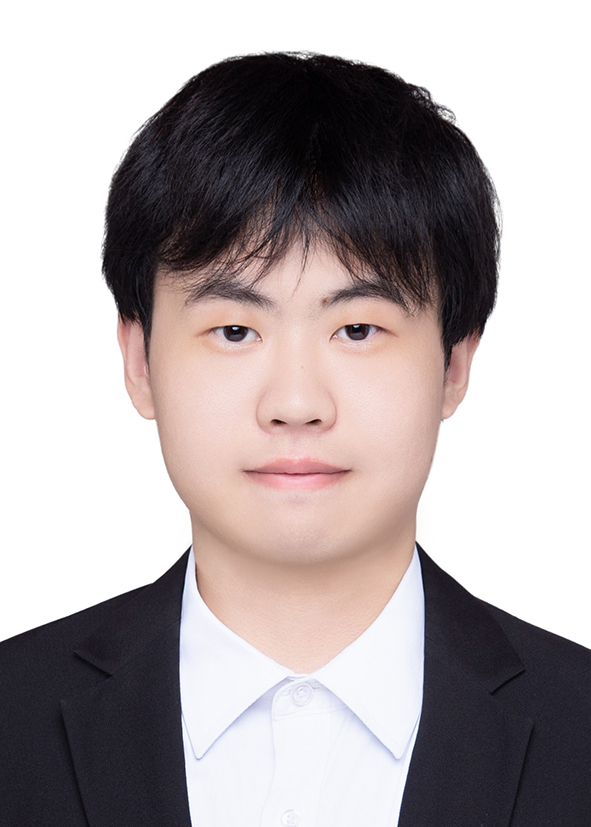}}]{Zhongzheng Mu} received the B.E. degree from Zhejiang University Of Science \& Technology and the M.S. degree from Zhejiang Gongshang University. 
His research interests include blockchain and smart contract security. 
\end{IEEEbiography}

\vspace{-10pt}

\begin{IEEEbiography}
[{\includegraphics[width=1in,height=1.25in,clip,keepaspectratio]{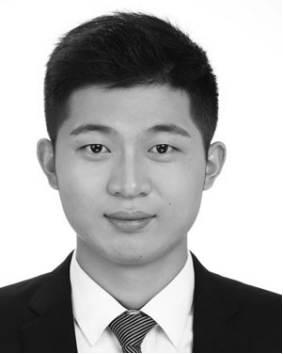}}]{Chengkun Wei} is currently a research professor of Zhejiang University, Hangzhou, China. 
He received the MS degree in computer science and technology from the Second Institute of China Aerospace Science, and the PhD degree in computer science and technology from Zhejiang University. 
His research interests include privacy protection and blockchain. 
\end{IEEEbiography}

\vspace{-10pt}

\begin{IEEEbiography}[{\includegraphics[width=1in,height=1.25in,clip,keepaspectratio]{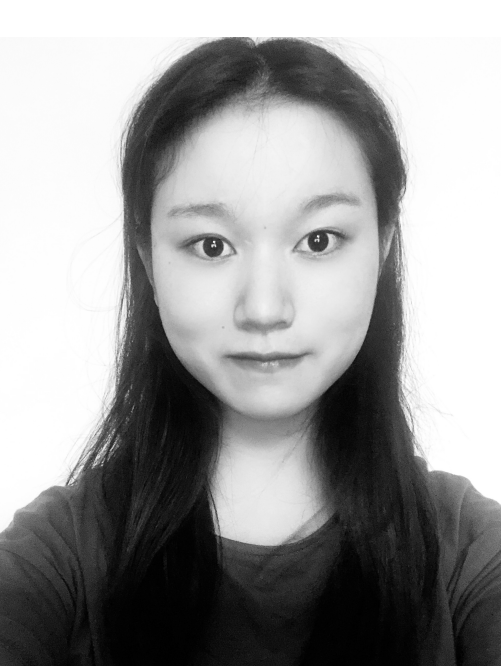}}]{Xiaojun Xu} 
received the B.E. degree from Taizhou University and the M.S. degree from Zhejiang Gongshang University. Her research interests include blockchain and smart contract security. 
\end{IEEEbiography}

\vspace{-10pt}

\begin{IEEEbiography}[{\includegraphics[width=1in,height=1.25in,clip,keepaspectratio]{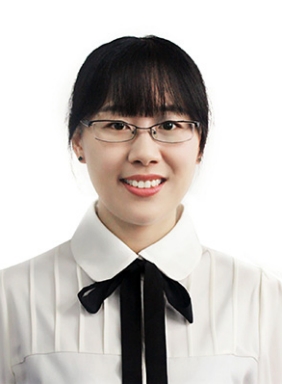}}]{Yingying Jiao} 
is currently pursuing the Ph.D. degree in the Department of Computer Science and Technology, Jilin University, China. 
She had been a research assistant with the School of Computing, National University of Singapore, Singapore. 
Her research interests include blockchain technology, deep learning, and image processing. 
\end{IEEEbiography}

\vspace{-10pt}

\begin{IEEEbiography}[{\includegraphics[width=1in,height=1.25in,clip,keepaspectratio]{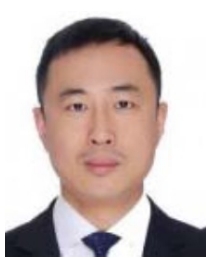}}]{Kui Ren} (IEEE Fellow, ACM Fellow) received the PhD degree from Worcester Polytechnic Institute. 
He is currently a professor and the dean with the College of Computer Science and Technology, Zhejiang University, where he also directs the Institute of Cyber Science and Technology. Before that, he was the SUNY Empire Innovation professor with the State University of New York at Buffalo. He has authored or coauthored extensively in peer-reviewed journals and conferences. His research interests include data security, IoT security, AI security, and privacy. 
\end{IEEEbiography}

\vspace{-10pt}

\end{document}